\documentclass[usenatbib]{mnras}

\usepackage{multirow}
\usepackage{rotating}
\usepackage{colortbl}
\usepackage{color}
\usepackage[fleqn]{amsmath}
\usepackage{amssymb}
\usepackage{amsfonts}
\usepackage{verbatim}
\usepackage{scalefnt}
\usepackage[percent]{overpic}
\usepackage[T1]{fontenc} 
\usepackage{aecompl} 
\usepackage{ulem}
\usepackage{cleveref}
\usepackage{todonotes}

\usepackage{float}

\def\lsim{\mathrel{\rlap{\lower 3pt \hbox{$\sim$}} \raise 2.0pt \hbox{$<$}}}
\def\gsim{\mathrel{\rlap{\lower 3pt \hbox{$\sim$}} \raise 2.0pt \hbox{$>$}}}

\newcommand{\comments}[1]{} 

\title[Dirty waveforms]{Dirty waveforms: multiband harmonic content of gas-embedded gravitational wave sources}

\author[L. Zwick et al.]{Lorenz Zwick,\thanks{E-mail: zwicklo@ics.uzh.ch} Andrea Derdzinski, Mudit Garg, Pedro~R. Capelo and Lucio Mayer
\\
Center for Theoretical Astrophysics and Cosmology, Institute for Computational Science, University of Zurich,\\ 
Winterthurerstrasse 190, CH-8057 Z{\"u}rich, Switzerland
}

\date{Accepted XXX. Received YYY; in original form ZZZ}

\pubyear{2022}

\begin{document}

\label{firstpage}

\pagerange{\pageref{firstpage}--\pageref{lastpage}}

\maketitle


\begin{abstract}
We analyse the effect of stochastic torque fluctuations on the orbital evolution and the gravitational wave (GW) emission of gas-embedded sources with intermediate and extreme mass ratios. We show that gas-driven fluctuations imprint additional harmonic content in the GWs of the binary system, which we dub \textit{dirty waveforms} (DWs). We find three interesting observational prospects for DWs, provided that torque fluctuations do indeed persist beyond the resolution limit of current hydrodynamical simulations. Firstly, DWs can produce a significant stochastic GW background, comparable to other GW noise sources. Secondly, the energy flux implied by the additional harmonics can cause a detectable secular phase shift in Laser Interferometer Space Antenna (LISA) sources, even if the net torque fluctuations vanish when averaged over orbital time-scales. Lastly, the DWs of moderate-redshift nHz supermassive binaries detectable by pulsar timing arrays (PTAs) could be detectable in the mHz range, producing a new type of PTA-LISA multiband gravitational source. Our results suggest that searching for DWs and their effects can potentially be a novel way to probe the heaviest of black holes and the physics of the accretion discs surrounding them. We find these results to be a further confirmation of the many exciting prospects of actively searching for environmental effects within the data stream of future GW detectors.
\end{abstract}



\section{Introduction} \label{sec:introduction}

The Laser Interferometer Space Antenna \citep[LISA;][]{LISAwhite2017,LISAwhite2020,Barack2019} is scheduled to fly in the early 2030s and it will allow for the detection of gravitational waves (GWs) in the mHz frequency band. Some of the most exciting sources of GWs for LISA are intermediate and extreme mass ratio inspirals. These consist of compact-object binary systems with a total mass of $10^4$ to $10^7$~M$_{\sun}$, where the secondary mass is a factor $\sim$$10^2$ to $\sim$$10^5$ smaller than the primary. These types of sources are expected to complete thousands of orbits in LISA's frequency band, and the resulting high signal-to-noise ratios (SNRs) make them a primary candidate to test general relativity \citep[GR; see, e.g.][]{Barack2007,Yunes2012,Wenbiao2017}. However, LISA sources are situated within noisy astrophysical environments and their evolution can therefore deviate from the pure vacuum predictions of GR. The detection of GWs relies on filtering the data with pre-produced waveform templates. If not accounted for, environmental effects introduce errors in the parameter fitting process, which could potentially ``spoil precision gravitational wave astrophysics'' \citep[][]{Barausse2014}. Clearly, quantifying and modelling the influence of the environment on GW inspirals is an essential exercise if one hopes to test GR in the strong-field regime.

However, it is also possible to reverse this logic and see deviations from the vacuum predictions as an opportunity to \textit{measure} properties of the environment \citep[see, e.g.][]{Ford2019,Cardoso2020}. In this context, the type of sources that seem most promising are compact objects embedded in a thin accretion disc surrounding a primary supermassive black hole (SMBH). A GW source of this kind can be created by several different formation channels. Star formation processes within the accretion disc itself can produce stellar-mass compact remnants, which subsequently migrate towards the primary \citep[][]{Goodman2004,Levin2007}. Alternatively, a population of stellar-mass compact objects can be dragged into alignment with the accretion disc by repeatedly crossing through it over the course of several orbits \citep[see, e.g.][]{Artymowicz1993,Rauch1995,Fabj2020}. Recent work has suggested that the latter type of gas-embedded inspiral might dominate the event rates of detectable extreme mass-ratio sources over more established processes such as dynamical relaxation \citep[][]{wetEMRI2021}, and that the presence of black holes (BHs) within accretion discs might be crucial for their stability \citep[][]{Gilbaum2021}. Another channel that can produce gas-embedded inspirals are galaxy mergers. After a merger event, two SMBHs can be brought to sub-pc separations by a variety of processes, including dynamical friction, gas drag, and gravitational interactions with clumps and galactic bars \citep[see, e.g.][]{Mayer2013,Tamburello2017,SouzaLIma2020,Bortolas2020}. These sources can have total masses ranging from $10^{5}$~M$_{\sun}$ all the way up to $10^{10}$~M$_{\sun}$. On the more massive end, they would produce very loud GWs, but often at frequencies lower than LISA's preferred sensitivity band.

In general, we expect a compact object embedded in a thin gaseous disc to be subject to torques. These are caused by the response of the disc to a massive perturber, in a way that is analogous to the well studied planetary migration. Simple analytical estimates suggest that gas torques can dephase the GW signal of an inspiral to detectable amounts, provided that the density of the disc is high enough \citep[][]{Kocsis2011,Barausse2014}. More recent hydrodynamical simulations have confirmed these estimates \citep[][]{Derdzinski2019,Derdzinski2021}, solidifying the case that the influence of the accretion disc will be present in the GW signal of gas-embedded LISA sources. If isolated from the vacuum prediction, the dephasing could be used to estimate disc properties such as density and temperature.

The aforementioned literature has focused primarily on the linear torque regimes, that can be approximately described with the simple formulae provided in the seminal works of \cite{Ward1997} and \cite{Tanaka2002}. However, gas torques exerted on a secondary body can deviate from linear estimates or exhibit nonlinear behaviours, particularly in the gap-opening regime \citep{Duffel2015}. 
In this paper, we focus on the strong stochastic fluctuations which are commonly observed in simulations (see Section~\ref{sec:Methods:Stochastic} for a more thorough introduction) on top of the linear effect. We investigate their imprint on the GW emission of a gas-embedded source, and find several interesting consequences relevant for future GW detectors. While we focus on stochastic gas torques as a physically motivated example, our results are more general, and can be applied to any process that is described as a small constant effect with high-amplitude fluctuations that average out to zero over sufficiently long time-scales.

The structure of this paper is as follows: in Section~\ref{sec:Methods}, we describe our simplified accretion disc model, and our assumptions for both linear and stochastic torques regimes. We reproduce and briefly discuss some known results for binaries subject to linear torques. In Section~\ref{sec:DirtyWaveforms}, we derive the effect of stochastic torque fluctuations on the motion of a binary of compact objects. We show that torque fluctuations leave an imprint in the binary's GW frequency spectrum, and how they can produce a secular dephasing in addition to the linear torque effect. In Section~\ref{sec:Detectability}, we discuss the detectability of the additional harmonics, both for individual sources as well as a GW background from a population of gas-embedded SMBH binaries. We also discuss the relevance of the secular dephasing for typical LISA sources. Our findings suggest that all of the effects mentioned above will be relevant for mHz GW detectors, provided that torque fluctuations persist beyond the resolution limit of current hydrodynamical simulations. Finally, we summarise our findings, discuss several caveats, and present some concluding remarks in Sections~\ref{sec:Discussion} and~\ref{sec:Discussion:conclusion}.

\section{GW Inspirals in Thin Discs}\label{sec:Methods}

In this section, we present and discuss the quantities needed to describe the inspiral of a compact object in a thin disc. In Section~\ref{sec:Methods:disk}, we define a simple but general disc model, which we will use to derive several useful scaling relations and order-of-magnitude estimates. This simple model will be sufficient for the purpose of this paper, and can be used to extrapolate our results to more complex accretion discs. In Section~\ref{sec:Methods:Linear}, we briefly summarise the effect of linear torques on the inspiral of a GW source, reproduce some known results on the dephasing of the waveforms, and refer the reader to the relevant literature. In Section~\ref{sec:Methods:Stochastic}, we define our model for stochastic torque fluctuations and describe how it can be applied in the context of GW generation.

\subsection{Simple disc model}\label{sec:Methods:disk}

For the purposes of this paper, we adopt a thin, power-law disc model which resembles a simplified version of the well known $\alpha$-disc model of \cite{Shakura1973}. Our goal is to have the simplest possible radial scaling of the disc properties, while still retaining some basic features of more realistic models. One such feature is mass conservation throughout the annuli of the disc, which puts a simple constraint on radial scalings. Assuming that the central SMBH accretes at a steady-state rate, we have
\begin{align}
     \dot{M}_{\rm{disc}}\sim 3 \pi \nu \Sigma \sim \textit{constant,}
     \label{eq:masscons}
\end{align}

\noindent where $\Sigma$ is the surface density, $\nu=\alpha c_{\rm s}^2 / \Omega$ is the kinematic viscosity, $\alpha<1$ is the dimensionless viscosity parameter, $c_{\rm s}$ is the speed of sound, and $\Omega$ is the disc's Keplerian orbital velocity.  
We define a simple disc model given in terms of a density profile $\rho(r)$ and a speed-of-sound profile $c_{\rm s}(r)$:
\begin{align}
\label{eq:diskmodel}
  \rho(r) &= \rho_{0}\left( \frac{3 r_{\rm S}}{r}\right)^{3/2},\\
  \label{eq:diskmodelcs}
  c_{\rm s}(r) &= 2\times 10^6 \left(\frac{3 r_{\rm{S}}}{r} \right)^{1/2}\, \left[\rm{m}\,\rm{s}^{-1}\right],
\end{align}

\noindent where $r_{\rm S}$ denotes the Schwarzschild radius of the primary BH and $\rho_0$ is a free parameter that allows us to describe a variety of possible disc densities. Note that we scale the radial profiles with the value at the innermost stable circular orbit (ISCO) of a non-spinning BH. These power-law profiles satisfy the requirement of mass conservation (Eq.~\ref{eq:masscons}) and reproduce the results of more complete $\alpha$-disc models with sufficient accuracy within a range of tens to thousands of Schwarzschild radii for a density parameter $\rho_0$ of
\begin{align}
    \label{eq:diskdensity}
    \rho_0 \approx 7.8 \times 10^{-6} \left( \frac{10^4\,\rm{M}_{\odot}}{\alpha M_1}\right)^{7/10}\left(\frac{\beta}{\eta} \right)\left[\rm{g}\,\rm{cm}^{-3}\right].
\end{align}

Here $M_1$ is the mass of the primary BH, $\beta$ is the fraction of the accretion rate with respect to the \citet{Eddington1916} limit, and $\eta$ is the efficiency of energy conversion for infalling gas. The values of these three parameters are generally unconstrained, and are likely to vary from disc to disc. Unless stated otherwise, we assume that the primary is accreting at small fraction of the Eddington limit ($\beta\sim 0.1$), with an efficiency consistent with a slowly rotating BH ($\eta\sim 0.1$). For a primary BH of $10^6$ M$_{\sun}$ and a viscosity $\alpha = 0.01$, this gives us a density parameter of the order $\rho_0 \approx 10^{-5}$~g~cm$^{-3}$. Depending on the choice of $\alpha$, $\beta$, and $\eta$, the density can easily vary by an order of magnitude in both directions. Therefore, we keep $\rho_0$ itself as a free parameter, with which we can scale all subsequent results. 

The radial scaling of Eq.~\eqref{eq:diskmodel} with a density normalisation of the order of Eq.~\eqref{eq:diskdensity} yields typical densities of the order of $10^{-10}$~g~cm$^{-3}$ at separations of thousands of Schwarzschild radii, consistent with results from more sophisticated disc models \citep[see, e.g.][]{SirkoGoodman2003,Thompson2009} as well as extrapolations from observed disc masses \citep[see, e.g.][where such extrapolations are discussed]{Mestel1963,Medling2014,Zwick2021}. Moreover, the total enclosed mass $M_{\rm{disc}}$ of the disc within a given radius $r$ amounts to
\begin{align}
    \label{eq:encl}
   \frac{ M_{\rm{disc}}(r)}{M_{\odot}}\sim  10^{-3}  \left(\frac{r}{r_{\rm S}}\right)^{3/2} \left(\frac{M}{10^6 \, \rm{M}_{\odot}} \right)^{3}\left(\frac{\rho_0}{10^{-5}\, \rm{g}\,\rm{cm}^{-3}} \right),
\end{align}

\noindent which is a negligible fraction of the total mass of the binary system $M$, and in most cases much smaller than the mass of the secondary.

Note that most of the calculations in this paper only depend on the value of the local density. Therefore, the density normalisation $\rho_0$ and our specific choice of a radial scaling can be extrapolated to other disc models. As an example, to obtain a result for a different density profile with a scaling of $\rho= \rho_{n} (3 r_{\rm S}/r)^n$, one can replace $\rho_0$ with an adjusted value that would yield the same local density:
\begin{align}
    \rho_0 \to \rho_{n} \left(\frac{3 r_{\rm S}}{r}\right)^{n-3/2}.
\end{align}

Within the context of this paper, which investigates the size and detectability of a new type of environmental influence in the GW emission of gas-embedded sources, we believe that our simple disc model suffices as an order-of-magnitude estimate.

\subsection{Linear torques} \label{sec:Methods:Linear}

Objects embedded in a thin gaseous disc will be subject to global torques, which are caused by the response of the gas to their presence. Here, we wish to characterise the evolution of an inspiralling binary of compact objects, which is embedded on a prograde orbit in a thin accretion disc. We focus on binaries with small mass ratios $q \ll 1/25$, such that the gas interaction is dominated by angular momentum exchange with the secondary. In this limit, torques are likely to be well described by the two standard regimes know in the literature as Type I and Type II \citep{Ward1997}. The former is valid 
for the smallest mass ratios ($q\lesssim10^{-4}$), for which torques are well described by linear perturbation theory\footnote{Note that Type I torques can also be sensitive to disc thermodynamics \citep{Paardekooper2006,Paardekooper2010}, but in this work we assume the locally isothermal regime for which the linear estimates have been successfully tested with numerical studies by  \cite{Tanaka2002}.}\citep[][]{Goldreich1980}. The latter occurs when the secondary is large enough to open a gap (i.e. lower the gas density in its coorbital region), which typically occurs for larger mass ratios ($q\gtrsim10^{-3}$).  
In this case, the torques become weaker than in the Type I regime,
although nonlinearities develop in the gas flow across the gap that challenge simple analytical predictions  \citep{Duffell2014}. 
Historically, Type II torques are assumed to follow the rate of radial gas inflow due to the disc viscosity \citep[see, e.g.][]{Edgar2007}, which is the assumption we adopt in this work.

Over time-scales that are larger than a single orbit, the secular orbital evolution is determined by the average fluxes of energy $E$ and angular momentum $L$ that the binary exchanges with its environment. The additional effect of gas torques will appear alongside with the GW-induced fluxes,
\begin{align}
\label{eq:energyeq}
    \left<\dot{E}\right>&= \left<\dot{E}_{\rm{GW}}\right>+ \left<\dot{E}_{\rm{T}}\right>\\
    \left<\dot{L}\right>&= \left<\dot{L}_{\rm{GW}}\right> + \left<\dot{L}_{\rm{T}}\right>
\end{align}

\noindent where the subscripts $\rm{GW}$ and $\rm{T}$ denote the GW- and torque-induced effects respectively, and the brackets denote an orbital average. Whereas $\dot{L}_{\rm{T}}$ is simply the value of the gas torques, the average energy flux associated to a torque is
\begin{align}
    \dot{E}_{\rm{T}} = \dot{L}_{\rm{T}}\omega_{\rm{o}} + \mathcal{O}\left[ e^2\right],
\end{align}

\noindent where $\omega_{\rm{o}}$ is the orbital angular velocity and $e$ is the orbital eccentricity. Here we have assumed that the orbit is almost circular. Therefore, any force applied by the torques is close to parallel to the tangential velocity.

From this point on, we assume that the standard formulae for Type I and Type II migration torques commonly used in the literature are a valid order-of-magnitude estimate for a compact object embedded in an accretion disc. In terms of the local density and the speed of sound, the formulae read \citep[see, e.g.][]{Ward1997,Tanaka2002}
\begin{align}
    \label{eq:typeI}
    \dot{L}_{\rm{I}}^{\rm{lin}}&=-q^2\frac{ \rho(a) \sqrt{a^3 G^3 M^3}}{4 \pi ^2  c_{\rm s}(a)},\\
    \dot{L}_{\rm{II}}^{\rm{lin}}&=-\alpha \frac{6 \pi a^{7/2}  c_{\rm s}(a)^3 \rho(a) }{ \sqrt{G M}},
    \label{eq:typeII}
\end{align}

\noindent where $a$ is the orbital separation (or semi-major axis) of the binary and $G$ is the gravitational constant. Note that the minus sign in front of both formulae signifies that the effect of the torque is to reduce the angular momentum of the binary. While this seems to be the standard case (for $q\lesssim 1/25$), simulations have shown that the sign of the torque can actually vary depending on the mass ratio due to non-linear effects \citep[see, e.g.][]{Duffel2015,Derdzinski2019,Derdzinski2021}. In such cases, the magnitude of the torque can still be approximated by Eqs~\eqref{eq:typeI} and \eqref{eq:typeII}, but with a reversed sign.

For any disc model in which the speed of sound scales as an inverse square root of the radius (Eq.~\ref{eq:diskmodelcs}), the ratio between Type I and Type II torques is constant for all separations. Since only the Type I formula scales with the mass ratio, we can find a critical mass ratio $q_{\rm{TI}=\rm{TII}}$ at which the two torque regimes have the same strength. By equating Eq.~\eqref{eq:typeI} and Eq.~\eqref{eq:typeII} we find
\begin{align}
    \label{eq:qcrit}
    q_{\rm{TI}=\rm{TII}} \sim 7 \times 10^{-4} \left(\frac{\alpha}{0.01} \right)^{1/2}.
\end{align}

This gives us a simple criterion to select which torque regime to apply to different mass ratios. Unless stated otherwise, \textit{we always choose the regime that yields the weaker torque,} which assures that our estimates are either appropriate or an underestimation of the actual torques which are likely to be applied on the binary. Note that this simple approach yields very similar results to the several gap-opening criteria used in the literature \citep[see, e.g.][in the context of compact object binaries]{Baruteau2011,Dittmnann2015}.

We can relate the orbit-averaged energy and angular momentum fluxes to the secular variations of the orbital parameters \citep[][]{Newton1687}:
\begin{align}
\label{eq:adot}
 \dot{a}&= \dot{a}_{\rm{GW}}+ 2\frac{\dot{L}_{\rm{T}}}{M q}\sqrt{\frac{a}{G M}} +\mathcal{O}\left[ e^2\right] \equiv \dot{a}_{\rm{GW}}+\dot{a}_{\rm{T}},\\
 \dot{e}&= \dot{e}_{\rm{GW}}- e\frac{ \dot{L}_{\rm{T}}}{4  M q}\sqrt{\frac{1}{G M}} + \mathcal{O}\left[ e^3\right] \equiv \dot{e}_{\rm{GW}}+\dot{e}_{\rm{T}},
\end{align}

\noindent where $\dot{a}_{\rm{GW}}$ and $\dot{e}_{\rm{GW}}$ are the GW-evolution terms \citep[][]{Peters1964}. Since in general the sign of the gas torque is negative, we can deduce a few qualitative results:

\begin{itemize}
    \item Linear torques facilitate the decay of the semimajor axis, thus speeding up the inspiral.
    \item Perfectly circular inspirals remain so.
    \item The efficiency of circularisation is reduced. Beyond a critical separation, linear torques can induce an increase of the eccentricity.
\end{itemize}

\begin{figure}
    \centering
    \includegraphics[scale=0.7]{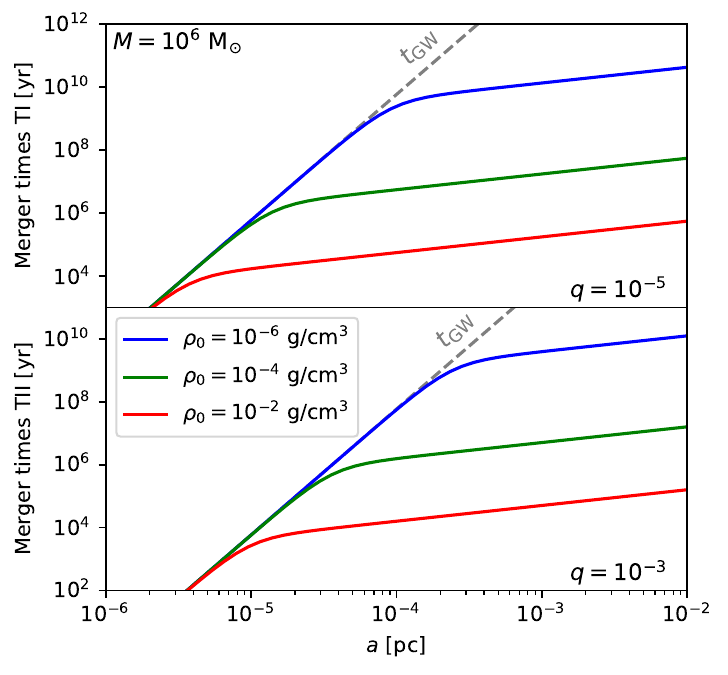}
    \caption{We show the results for the merger time-scales for a binary with two different mass ratios subject to gas torques within a thin disc with different density normalisations (shown by the different coloured lines). The results are obtained by solving Eq.~\eqref{eq:adot}. Clearly visible are the transitions between the GW-dominated phase at small separations and the gas-dominated phase at large separations. Increasing the normalisation density has the effect of shifting the transition radius closer to the primary BH.}
    \label{fig:timescales}
\end{figure}

While the first point is intuitive, the topic of eccentricity evolution of gas-embedded sources is rich and complex. The standard assumption is that gas effects will lead to a more efficient circularisation of the orbit \citep[see, e.g.][]{Barausse2014,Ford2019,Tanaka2004}. However, eccentricity pumping has been shown to occur for gap-opening secondaries \citep{Duffel2015}. In the context of LISA sources, significant eccentricities can also be excited by gas drag \citep[][]{Cardoso2021} or by circumbinary disc physics \citep[][]{Dorazio2021,Zrake2021}. In this work, we will assume that the binaries quickly circularise once they reach the GW-dominated phase. Further effects of the eccentricity will be discussed in Section~\ref{sec:Discussion}.

Eq.~\eqref{eq:adot} has two important consequences for GWs. Firstly, it affects the merger time-scales of binaries with a given separation, which has implications on the event rates and the residence times in different frequency bands \citep[see, e.g.][]{Haiman2009}. In Figure~\ref{fig:timescales}, we show the results for the merger time-scale obtained by integrating Eq.~\eqref{eq:adot} from different initial conditions to the ISCO of the primary BH. Clearly visible is the transition from the GW-dominated regime at small separations and the gas-dominated regime at large separations. Secondly, if the deviation from the vacuum inspiral implied by Eq.~\eqref{eq:adot} occurs within the frequency band of a GW detector, it will produce a potentially detectable phase drift. Since this is a crucial factor in discerning environmental effects within the LISA data stream, it is worth re-deriving some known results for linear torques.

\subsubsection{Relative energy flux and dephasing from smooth torques}

A simple way to quantify a deviation from the vacuum evolution of a binary is to compare the size of the environmentally caused energy flux to the lowest-order vacuum prediction, which is sourced by the quadrupole formula \citep[][]{Einstein1916}. We define the relative power\footnote{Note that the relative power $\mathcal{P}$ can also be easily compared with post-Newtonian corrections of the vacuum fluxes through equations of the form $\mathcal{P}\sim (r_{\rm S}/a)^n $, as done in \citet{Zwick2021}. This can give estimates for how precise waveform templates must be to resolve environmental effects.} $\mathcal{P}$ that is caused by an environmental effect with the equation
\begin{align}
    \left<\dot{E}\right>= \left<\dot{E}_{\rm q}\right>\left(1+ \mathcal{P} \right)
\end{align}

\noindent where $\dot{E}_{\rm q}$ is given by \citep[][]{Peters1963}
\begin{align}
\left< \dot{E}_{\rm q}\right>=  \frac{ G^4 M^5 q^2}{15 a^5 c^5} + \mathcal{O}\left[e^2\right]
\end{align}

\noindent and $c$ is the speed of light in vacuum. An additional energy flux causes the inspiral of the binary to proceed at a slightly different rate than expected. This in turn causes a phase shift of the GW signal from the vacuum waveform. For our purposes, we can find a simple relation between the relative power and the dephasing of the GW signal, $\delta \phi$, by manipulating the general dephasing equation for a source that merges according to Eq.~\eqref{eq:energyeq} \citep[see, e.g.][]{Kocsis2011,Barausse2014}:
\begin{align}
    \delta \phi = \phi_{\rm{vac}} - \phi \approx \int \omega_{\rm o}(a) \frac{\dot{a}_{\rm{T}}}{\dot{a}_{\rm{GW}}^2} {\rm d}a,
\end{align}

\noindent where $\phi_{\rm{vac}}$ and $\phi$ are the total orbital phase of the source in a vacuum or in a gaseous medium, respectively. Here we also assumed that we are in the GW-dominated regime of evolution. We can switch the integration from orbital separation to time, and use the fact that $\dot{a}$ is proportional to $\dot{E}$ to find:
\begin{align}
    \label{eq:dephasingintegral}
    \delta \phi \approx \int \omega_{\rm o} \frac{\dot{E}_{\rm{T}}}{\dot{E}_{\rm{GW}}} {\rm d}t \approx  \int\omega_{\rm o} \mathcal{P} {\rm d}t \sim \omega_{\rm o} \mathcal{P} T_{\rm{obs}},
\end{align}

\noindent where in the last step we assumed a monochromatic source and $T_{\rm{obs}}$ is the observation time. To relate Eq.~\eqref{eq:dephasingintegral} to a LISA observable, we redshift the source frequency, $f_{\rm z} = f(1+z)$, where $f$ is the observed frequency of the GW, obtaining
\begin{align}
    \label{eq:depheq}
    \frac{\delta \phi}{2 \pi} \approx  \frac{1}{2}f_{\rm z} \mathcal{P} T_{\rm{obs}}.
\end{align}

With the use of this formula, valid for monochromatic sources within the GW-dominated phase of the evolution, we can find very simple scaling relations for the expected dephasing of a GW source that is observed by LISA for a time $T_{\rm obs}$. For our disc model and our estimates for Type I (TI) and Type II (TII) torques, we find the following results:
\begin{align}
    \label{eq:dephasingTIlin}
     \delta \phi_{\rm{TI}} &\sim 0.8 \left(\frac{10^{-4} \, \rm{Hz}}{f_{\rm z}} \right)^{4/3} \left(\frac{10^6 \, \rm{M}_{\odot}}{M} \right)^{1/3} \nonumber \\
     &\times \left(\frac{\rho_0}{10^{-5}\rm{g}\,\rm{cm}^{-3}}\right) \left(\frac{T_{\rm{obs}}}{\rm{yr}}\right), \\
     \label{eq:dephasingTIIlin}
     \delta \phi_{\rm{TII}} &\sim 6.2 \times \frac{\alpha 10^{-5}}{q^2}  \left(\frac{10^{-4} \, \rm{Hz}}{f_{\rm z}} \right)^{4/3} \left(\frac{10^6 \, \rm{M}_{\odot}}{M} \right)^{1/3} \nonumber \\
     &\times \left(\frac{\rho_0}{10^{-5}\rm{g}\,\rm{cm}^{-3}}\right) \left(\frac{T_{\rm{obs}}}{\rm{yr}}\right).
\end{align}

As expected, the dephasing increases linearly with observation time and the density parameter, while it decreases for sources with higher mass or at higher redshift, where we observe the inspiral at later stages.
Note that while these results are derived assuming monochromatic sources, they can also serve as order-of-magnitude estimates for the dephasing of chirping sources. This is because most of the dephasing accumulates at the initial separation, where the source completes most of its cycles and the relative strength of the gas effects is large.

The phase of a GW signal can be reconstructed with an accuracy of approximately $1/\rm{SNR}$ \citep[see, e.g.][]{Glampedakis2006,Katz2021}. Typically, LISA sources will have SNRs of a few to a few hundreds \citep[][]{LISAwhite2017,LISAwhite2020}, meaning that phase shifts of fractional order are likely to be detectable.\footnote{Here we neglect the problem of degeneracies between environmental effects and chirp mass. We discuss this problem in more detail in Section~\ref{sec:Discussion:caveats}.} The estimates given in Eqs~\eqref{eq:dephasingTIlin} and \eqref{eq:dephasingTIIlin}, while specific to our disc model, reproduce the magnitude of known results \citep[see, e.g.][]{Barausse2014,Derdzinski2021} and indicate that the dephasing of GW sources due to linear gas torques should indeed be a noticeable effect for LISA sources. We will use them as a benchmark to compare with the secular dephasing caused by stochastic torques in Section~\ref{sec:Detectability:individual}. 

\subsection{Torque fluctuations}\label{sec:Methods:Stochastic}

\subsubsection{Disc-driven versus perturber-driven variability}

We now consider the evolution of a binary due to stochastic torques, or time-variable torques that deviate from the smooth, linear estimates. In this section, we introduce two types of torque fluctuations based on the physical origin. Each predicts a characteristic spectrum of variability, which we describe with a generalized model. 
\newline

The first, and likely the most ubiquitous case in gas-embedded systems, are {\it disc-driven} fluctuations caused by instabilities in accretion discs. While the linear torque theory (Eqs~\ref{eq:typeI} and \ref{eq:typeII}) assumes a laminar gas flow,  accretion discs are expected to be turbulent. Indeed in the case of active galactic nucleus (AGN) discs, turbulence driven by magneto-rotational instability (MRI, \citealt{BalbusHawley1991}) is considered the main driver of angular momentum transport. Similarly protoplanetary discs, although less ionized, also exhibit turbulence due to MRI or other hydrodynamical instabilities \citep{Klahr2018}. A migrating perturber in a turbulent disc will experience torque fluctuations as it encounters density enhancements and rarefactions. Several planetary migration works confirm that turbulence produces a stochastic torque component in addition to the net, linear (Type I) torque evolution \citep{Nelson2005,Yang2009}. In cases of low perturber-to-disc mass ratios, stochastic migration can take over the evolution entirely \citep{Johnsonetal2006}.

A second source of torque fluctuations occurs due to asymmetries in the gas flow near a sufficiently massive perturber. These fluctuations are observed in recent high-resolution hydrodynamical simulations of intermediate mass-ratio inspirals ($q\! =\! 10^{-3}$) embedded in 2D, laminar accretion discs with Mach numbers greater than $\mathcal{M}\ge 20$, equivalent to disc aspect ratios $h/r\le 1/20$ \citep{Derdzinski2021}. Unlike the disc-driven case, these {\it perturber-driven} fluctuations arise from the gas flow within the gravitational influence radius of the secondary. Notably, the frequency and amplitude of the fluctuations increase with higher Mach number and with resolution, which indicates a physical origin for high-amplitude, super-orbital torque oscillations in near-Eddington AGN discs. Many physical processes could cause fluctuations that take place well within the influence radius of the secondary, examples being stochastic accretion \citep[][]{Kelly2011}, the tidal influence of field stars \citep[][]{Gupta2021}, hydromagnetic effects \citep[as mentioned in e.g.][]{Mayer2006},  friction through a turbulent medium, or even simply small-scale gas dynamics. A further complication is due to the fact that there will likely be significant differences between the fluctuation spectra of retrograde versus prograde orbiters. In the case of the former, the smooth torque component will primarily be caused by supersonic drag \citep[see, e.g.][]{ostriker}, and therefore be much weaker in magnitude with respect to global torques. A retrograde orbiter would encounter inhomogeneities at much higher cadence, possibly leading to higher-frequency disc-driven fluctuations. We also expect differences in the small-scale flow close to the perturber, although it is harder to speculate how these would affect the torque in the perturber-driven case.

Such torque variability (and its dependency on system properties) deserves more exploration. If confirmed, it will produce several interesting observational signatures in the GWs of gas-embedded binaries, as we demonstrate in Section~\ref{sec:DirtyWaveforms}.

\begin{figure}
    \includegraphics[width=0.48\textwidth]{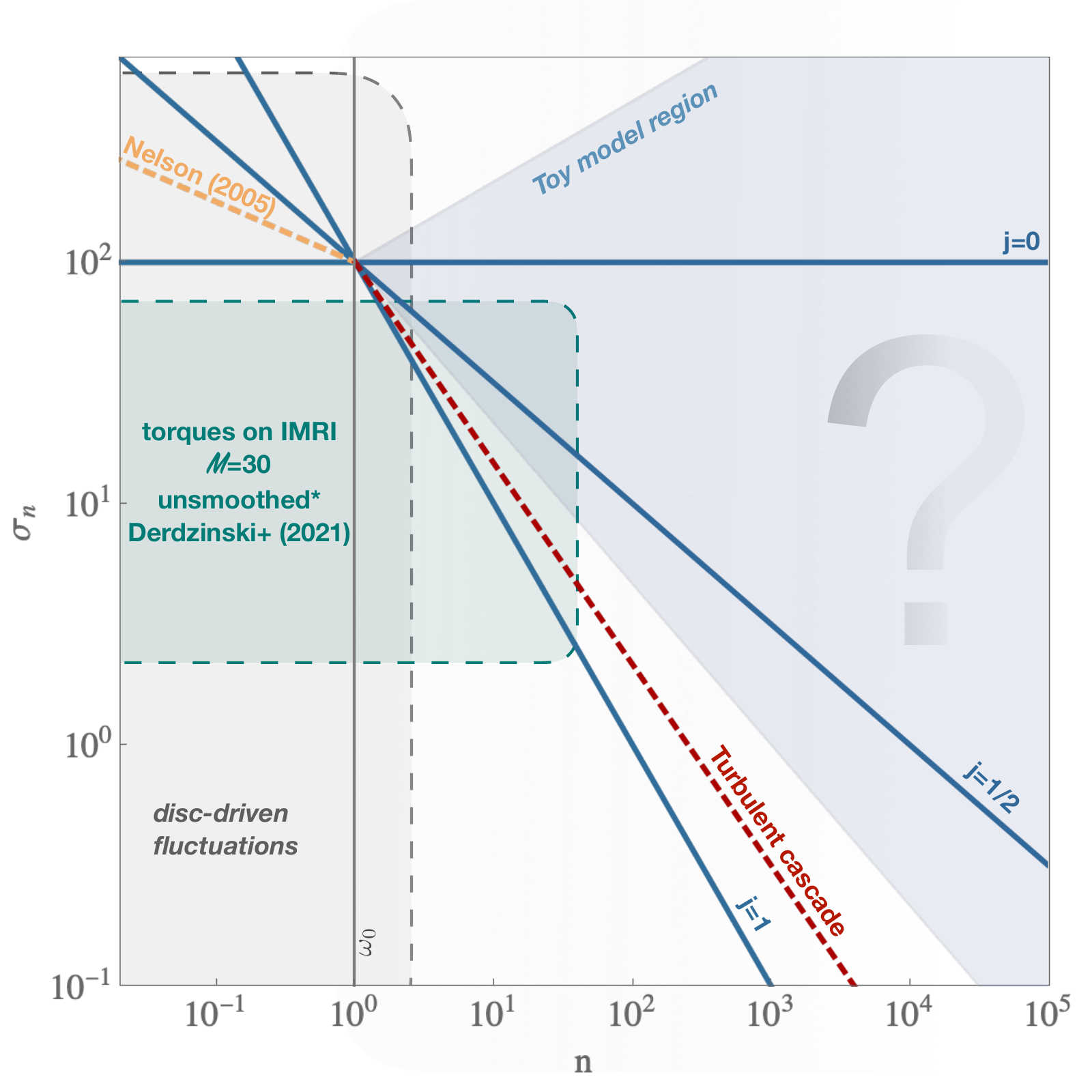}
    \caption{Sketched parameter space of stochastic torque amplitude $\sigma_n$ versus harmonic number $n$, for disc-driven or perturber-driven fluctuations. Blue lines delineate our fiducial models (Eq.~\ref{eq:flucamplitude}) for three different fluctuation spectral indices $j$. We highlight a cone shaped region (light blue) suggested by a simple toy model explained in the main text. We also show a red, dashed line motivated by the turbulent-cascade mechanism, and an approximate scaling for sub-orbital fluctuations \citep[orange dashed line adapted from][where we increased the amplitude by a factor $\sim 2$ to fit our normalisation]{Nelson2005}. We stress that our toy-model estimates should be only taken as loose suggestions for the range of spectral indices in the super-orbital region of the plot. 
    The green shaded region is based on simulation data from \citet{Derdzinski2021} of 
    a $q=10^{-3}$ intermediate mass-ratio inspiral embedded in discs with Mach number $\mathcal{M}\sim30$.}
    \label{fig:paramspace}
\end{figure}

\subsubsection{A general stochastic torque model}

A fully general torque fluctuation curve acting on a binary of compact objects can by expressed as a Fourier series over a period of $l$ orbits:
\begin{align}
\label{eq:stochTparam}
\dot{L}_{\rm T}^{\rm{stoch}}(t) &= \sum_{n=1}^{\infty} A_{n}^{\rm{T}} \cos(n \omega_{\rm o}t/l) + B_{n}^{\rm{T}} \sin(n \omega_{\rm o}t/l),
\end{align}

\noindent where $n$ is the harmonic number and the coefficients $A_{n}^{\rm{T}}$
and $B_{n}^{\rm{T}}$ have units of torque. Note that the net value of such a torque fluctuation curve vanishes when averaged over $l$ orbits. This is a crucial point for our analysis, since we want to make sure to disentangle the effect of torque fluctuations from the effect of a small constant torque.

The magnitude of the coefficients should in principle be read off from hydrodynamical simulations. However, due to the stochastic nature of gas flow and turbulence, it is impossible to say whether two separate realisations of $l$ orbits would necessarily yield an identical torque fluctuation curve. A more appropriate description is to say that the coefficients $A_{n}^{\rm{T}}$ and $B_{n}^{\rm{T}}$ are indeed \textit{random variables}, selected every $l$ orbits from distributions with a given variance. In the case of planetary migration, such distributions have been found to be approximately Gaussian \citep[][]{Nelson2005}. In the remainder of this work, we interpret the torque fluctuation coefficients as being the ``typical pick'' from these distributions, which would naturally be of order of their standard deviation.

However, as we have mentioned before, simulations have yet to achieve the appropriate Mach numbers and resolutions to fully capture gas physics around small secondary BHs. Therefore, we are forced to assume a range of simple models for the typical amplitudes of the torque fluctuations. First off, we rescale the coefficients with the linear torque values, as is often done in the literature \citep[see e.g.][]{Tanaka2002,Derdzinski2019,Derdzinski2021,Dorazio2021}:
\begin{align}
    \label{eq:fluctscaling1}
    A_{n}^{\rm{T}} &\sim   \sigma_{n}^{\rm{A}} \dot{L}_{\rm{lin}}, \\
    \label{eq:fluctscaling2}
    B_{n}^{\rm{T}} &\sim   \sigma_{n}^{\rm{B}} \dot{L}_{\rm{lin}}, 
\end{align}

\noindent where $\sigma_{n}^{\rm{A}}$ and $\sigma_{n}^{\rm{B}}$ are dimensionless, and we assume no preference for the phase of the fluctuations, which implies $\sigma_{n}^{\rm{A}} \sim \sigma_{n}^{\rm{B}}$. In the context of this work, we define a series of power-law models for the typical fluctuation amplitudes,
\begin{align}
    \label{eq:flucamplitude}
    \sigma_{n} = \sigma \left( \frac{l}{n}\right)^j,
\end{align}

\noindent where $\sigma$ is the size of a fluctuation at the orbital frequency and the spectral index $j$ describes how quickly higher harmonics decay. It is important to note here that some physical processes could produce fluctuations at very specific frequencies, resulting in a spectrum with distinct peaks or a different shape altogether. In this sense, our simple power-law model is an arbitrary choice, convenient to calculate results for a large region of parameter space. We believe that this suffices for the purposes of this work, and our calculations can be easily extended to arbitrary fluctuation power spectra.

State-of-the-art simulations for gas-embedded sources with $q \sim 10^{-3}$ suggest that the typical size of $\sigma$ could be of order $10$ for Mach numbers $\mathcal{M} \gsim 20$ and of order $10^2$ for $\mathcal{M} \gsim 30$ \citep[based on unsmoothed results from][]{Derdzinski2021}, and results from planetary migration show a similar magnitude \citep[see, e.g.][]{Nelson2005}. AGN continuum emission observations indicate Mach numbers of order $\mathcal{M}\gtrsim 100$ \citep{Krolik1999}, which suggests that strong fluctuations are likely in realistic discs. Therefore, we take $\sigma \sim 10^2$ to be a reasonable estimate. The precise value of $\sigma$ is actually not crucial, since the physical size of the fluctuations can still vary by order of magnitudes scales depending on the disc model and the choice of Eddington ratio and accretion efficiency (see Eq.~\ref{eq:diskdensity} and further discussion in Section~\ref{sec:Discussion}). The parameter $j$ is yet to be constrained for super-orbital fluctuations\footnote{As a note, results in \cite{Nelson2005} suggest that the exponent $j$ is roughly $1/4$ for sub-orbital fluctuations in proto-planetary discs. Though interesting, we do not expect this scaling to extend further into to the perturber-driven regime.} due to resolution limits, smoothing constraints, and Mach number limitations of current simulations.

A few simple toy models suggest that $j$ will likely be in the region of $|j| \lsim 1$. As an example, turbulence can transport energy down to small scales and the power spectrum of the resulting ``cascade'' is known to scale with an exponent of $5/3$ in the Kolmogorov model \citep[which extends, with some modifications, to the supersonic, compressible regimes;][]{Kritsuk}. If a constant fraction of this energy is transmitted into the motion of the secondary, we expect the amplitude of the motion to scale as the square root of the energy, yielding $j=5/6$. In another toy model, fluctuations can be thought of as being sourced by the total gas mass $m_{\rm f}$ present within spheres of different radius $r_{\rm f}$ around the secondary. In the case of maximum asymmetry, the total mass is concentrated in a point at radius $r_{\rm f}$. The frequency of the fluctuation would then correspond to the orbital frequency at $r_{\rm f}$. In this model, the gravitational force acting on the secondary scales as $\sim m_{\rm f}/r_{\rm f}^2$, meaning that the power spectrum depends on the small-scale density profile $\rho_{\rm f}$ around the secondary BH. If the latter follows a power law, i.e. $\rho_{\rm f}\sim r_{\rm f}^{-\gamma}$, we find that $j=2(1-\gamma)/3$. Note that the limiting cases of $\gamma=0$ (homogeneous density) and $\gamma = 3/2$ (small-scale $\alpha$-disc around the secondary) correspond to $j = 2/3$ and $-1/3$, respectively, which is also indicative of a $j$ parameter of order $|j| \lsim 1$.

These simple toy models can only be used as a guideline to somewhat constrain the possible range of values for the spectral index $j$. Indeed, the question of the fluctuation power spectrum can only be resolved with sufficiently accurate numerical simulations, which have to include all physical phenomena of importance. Therefore, in this paper we focus on three arbitrary but reasonable values of the exponent $j$ ($0$, $1/2$, and $1$), which are shown in Figure~\ref{fig:paramspace} along with our toy-model estimates. With these three choices, we aim to sample a large region of the unknown parameter space of the perturber-driven fluctuation regime.

While this model is simplistic, note that we are not assuming that any given realisation of a torque fluctuation curve will follow such a simple power law. We interpret the amplitudes as standard deviations, meaning that any single torque curve is completely unconstrained. Our assumptions only constrain the fluctuation amplitudes in a statistical sense over many realisations, as justified by the Central Limit Theorem. In general, we expect $\sigma$ and $j$ to be functions of the disc parameters, the binary parameters and other details of the processes that are actually sourcing the fluctuations. Our estimates in this paper are speculative by necessity, directed by the limited amount of results on super-orbital torque fluctuations on binaries with small mass ratios. We are looking forward to revisiting them as more precise estimates emerge from future hydrodynamical simulations.

To complete our fluctuation model, we have to assure that high-frequency fluctuations are still physical, and that sums over all values of $n$ converge. Therefore, we do not allow fluctuations with frequencies higher than a maximum value given by the limiting case of a gas overdensity orbiting at the ISCO of the secondary. The maximum harmonic is found by dividing the latter with the orbital frequency of the binary itself, and is given by the following equation:
\begin{align}
    \label{eq:nmax}
      n_{\rm{max}} &= q^{-1} \left(\frac{a}{r_{\rm S}}\right)^{3/2}\\
    &\approx 2.3 \times 10^2 q^{-1}\left( \frac{10^{-4} \, \rm{Hz}}{f_{\rm{z}}} \right)\left(\frac{10^6 \, \rm{M}_{\odot}}{M} \right)
\end{align}
After having defined our simple models for the fluctuation amplitudes, we can finally describe their effect on the orbit of an inspiralling GW source.

\section{Stochastic Torque imprint on a GW source}\label{sec:DirtyWaveforms}

In this section, we show how the orbit of a gas-embedded inspiral responds to torque fluctuations, and how these leave imprints in the GW. In Section~\ref{sec:DirtyWaveforms:buffet}, we find a perturbative parametrization of the orbital trajectory for a source that is buffeted by stochastic torques. In Section~\ref{sec:DirtyWaveforms:harmonics}, we apply the quadrupole formalism to the buffeted orbit, and find that the GWs carry additional harmonic content at linear order in the perturbation. In Section~\ref{sec:DirtyWaveforms:flux}, we show that the additional harmonic content implies a secular energy flux at quadratic order in the perturbation.

\subsection{Buffeting of the orbit}\label{sec:DirtyWaveforms:buffet}

\begin{figure}
    \includegraphics[width=0.47\textwidth]{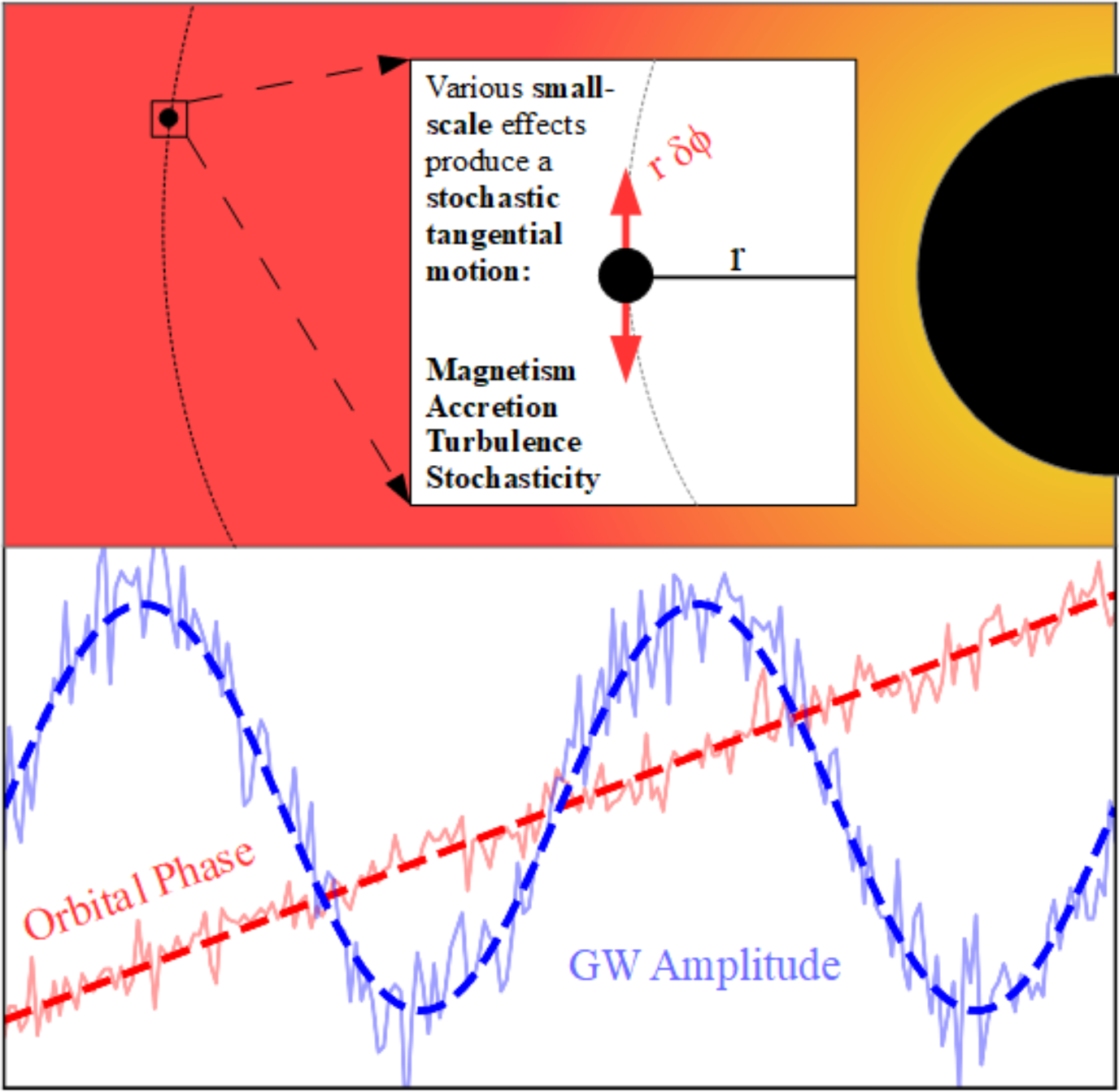}
    \caption{We show a simple visualisation of the effect of stochastic torques on the motion of a gas-embedded GW source. In the top panel, a compact object is perturbed by small-scale hydro-dynamical and magnetic effects within the accretion disc of the primary SMBH. At lowest order, this produces a stochastic, jittering motion in the tangential direction of the secondary's orbit. In the bottom panel, we show an exaggerated realisation of the DWs produced by such a system. Small high-frequency variations in the orbital phase (red) produce fluctuations in the GW amplitude (blue). We expect these perturbations to be several orders of magnitude smaller than the main monochromatic signal they ride on. The existence of DWs may have several interesting implications for planned mHz observatories, which we discuss in Section~\ref{sec:Detectability}.}
    \label{fig:nicepic}
\end{figure}

The physical effect of torque fluctuations on an orbiting body is to apply a force, which in turn causes an acceleration (see Figure~\ref{fig:nicepic} for a simple visualization of the phenomenon). One can imagine the secondary BH being constantly buffeted by the fluctuations, adding a small stochastic jitter on top of its circular motion. From here on, we assume Newtonian mechanics and gravity, two assumptions that are only valid at first order in $v^2/c^2$, where $v$ is the orbital velocity. However, we do not expect relativistic corrections to play a significant role for the results of this paper, as they would only couple to them as perturbative effects of even higher order.

We parametrize a small generic deviation from a circular, planar Keplerian orbit by means of a Fourier series:

\begin{align}
    \label{eq:rparam}
    r(t) &= a + a \delta \sum_{n=1}^{\infty} A_{n}^{r} \cos(n \omega_{\rm o} t/l) + B_{n}^{r} \sin(n \omega_{\rm o} t/l),\\
    \label{eq:phiparam}
    \phi(t)&= \omega_{\rm o} t + \delta \sum_{n=1}^{\infty} A_{n}^{\phi} \cos(n \omega_{\rm o} t/l) + B_{n}^{\phi} \sin(n \omega_{\rm o} t/l),
\end{align}

\noindent where $r(t)$ is the orbital separation, $\phi(t)$ is the orbit's phase, and the small dimensionless parameter $\delta$ is used to keep track of the size of the perturbation. Note that we neglect the $n=0$ contribution, since a small constant radial displacement can be reabsorbed in the definition of $a$, and a constant phase shift can be neglected by redefining the initial conditions of the parametrization. This assures that the average values of $r$ and $\phi$ over $l$ orbits are identical to the unperturbed circular case. We also assume that the torques are applied parallel to the angular momentum vector of the binary. In other words, we do not allow for the orbit to be buffeted out of alignment with the disc, reducing the problem to a 2D plane. We expect this to be a reasonable assumption for thin discs. Furthermore, a small change in azimuthal alignment does not influence the quadrupole moment of the binary (at lowest order), and GW generation is therefore unaffected.\footnote{Note that such a shift in alignment does indeed affect the GW polarisation amplitudes as seen by a given GW detector, which might also produce interesting observational signatures.}

To establish a connection between the torque fluctuation and the orbital parametrization, we take the Newtonian definition of torque, insert Eqs~\eqref{eq:rparam} and \eqref{eq:phiparam} and expand for small perturbations:
\begin{align}
    \label{eq:torquepert}
    \dot{L}_{\rm T}^{\rm{stoch}} &= \frac{{\rm d}}{{\rm d}t}\left[M q r(t)^2 \dot{\phi}(t)\right] \nonumber\\
    &= - \delta\sum_{n=1}^{\infty} \frac{a^2 n^2 M q \omega_{\rm o}^2}{l^2} \left( A_{n}^{\phi} \cos(n  \omega_{\rm o} t/l) + 
   B_{n}^{\phi} \sin(n \omega_{\rm o} t/l) \right) \nonumber \\
   &- \delta^2\sum_{n=1}^{\infty} \frac{a^2 n^2 M q \omega_{\rm o}^2}{l^2}  \left( A_{n}^{r} A_{n}^{\phi}  - B_{n}^{r}  B_{n}^{\phi}\right)\cos(2 n \omega_{\rm o}t/l) \nonumber\\
   &- \delta^2\sum_{n=1}^{\infty}\frac{a^2 n^2 M q \omega_{\rm o}^2}{l^2}  \left( A_{n}^{r} B_{n}^{\phi}  + B_{n}^{r}  A_{n}^{\phi}\right)\sin(2 n \omega_{\rm o}t/l) \nonumber \\ &+ \mathcal{O}[\delta^3].
\end{align}

From this equation, we can see that, at linear order, the only contribution of a torque fluctuation is to excite a corresponding perturbation of the phase of the orbit. At quadratic order, torque fluctuations can cause both phase and radial perturbations. Since we expect these perturbations to be small, we are content with solving Eq.~\eqref{eq:torquepert} at linear order, and find a simple connection between the torque fluctuation coefficients and the orbital perturbation coefficients:
\begin{align}
    \label{eq:coeffeq1}
    A^{\phi}_{n} &\approx A^{\tau}_{n}\frac{l^2}{a^2 M q n^2\omega_{o}^2},\\
    \label{eq:coeffeq2}
     B^{\phi}_{n} &\approx B^{\tau}_{n}\frac{l^2}{a^2 M q n^2\omega_{o}^2},\\
     A^{r}_{n} &\sim B^{r}_{n} \sim 0.
\end{align}

With the aid of Eqs~\eqref{eq:coeffeq1} and \eqref{eq:coeffeq2}, we can relate a torque fluctuation curve to the effect that it has on the secondary's orbit. Note that the actual physical displacement of the secondary is very small. If we insert our reference disc model (Eqs~\ref{eq:diskmodel} and \ref{eq:diskmodelcs}) and the estimate for Type I torques, we find
\begin{align}
   \frac{ a A_{\phi} }{q r_{\rm S}}&\sim 10^{-5} \sigma_{n} \left(\frac{10^{-4} \, \rm{Hz}}{f_{\rm z}} \right)^{5/3}\nonumber \\ & \times \left(\frac{M}{10^6 \, \rm{M}_{\odot}} \right)^{1/3}\left(\frac{\rho_0}{10^{-5}\, \rm{g}\,\rm{cm}^{-3}} \right),
\end{align}

\noindent where we normalised the displacement with the Schwarzschild radius of the secondary. The estimate for Type II torques is similar or smaller.

\subsection{Harmonic content of gravitational waves}\label{sec:DirtyWaveforms:harmonics}

Since there is a unique correspondence between the waveform and the orbit of its source, we expect to find a trace of the buffeting caused by stochastic torque fluctuations in the harmonic content of a binary's GWs. We use the standard formalism to produce the polarisation amplitudes at quadrupolar order:
\begin{align}
    h_{ij}= \frac{2G}{c^4 D_{\rm L}} \Ddot{Q}^{\rm{tot}}_{ij},
\end{align}

\noindent where $D_{\rm L}$ is the luminosity distance and $Q_{ij}^{\rm{tot}}$ is the total mass quadrupole tensor. We construct the perturbed binary's quadrupole tensor $Q_{ij}^{\rm{bin}}$ from our orbital parametrization given in Eqs~\eqref{eq:rparam} and \eqref{eq:phiparam}:
\begin{align}
     Q_{ij}^{\rm{bin}} = M q r_i(t)r_j(t),
\end{align}

\noindent where $r_k$ is equal to the projection $\mathbf{r}\cdot \mathbf{\rm{\mathbf{e}}}_{\mathbf{k}}$. Regarding the total mass quadrupole, we only keep the binary's contribution and neglect the presence of the gas, an assumption that is justified by the small enclosed disc mass within the separation $r(t)$ (see Eq.~\ref{eq:encl}):
\begin{align}
    Q_{ij}^{\rm{tot}} = M q r_i(t)r_j(t) +  Q_{ij}^{\rm{gas}} \approx  M q r_i(t)r_j(t).
\end{align}

After being projected into the transverse traceless gauge, the strain tensor reduces to two independent components $h_{+}$ and $h_{\times}$, whose amplitudes depend on the exact orientation of the source on the sky. In the context of this work, we are interested in the size and detectability of an effect from a general gas-embedded source, rather than an exact polarisation prediction. Therefore, we perform a sky average of the polarisation amplitudes. For an unperturbed circular source, the (sky-averaged) GW polarisation amplitudes scale with a typical strain, $\lvert h_{+} \rvert \sim \lvert h_{\times} \rvert \sim h_0$, given by the following equation:
\begin{align}
&h_0 = \frac{4 a^2 G M q \omega_{\rm o}^2}{c^4 D_{\rm L}}.
\end{align}

We find that a source that is perturbed according to Eqs~\eqref{eq:rparam} and \eqref{eq:phiparam} picks up additional harmonics contributions on top of the unperturbed signal. We denote the perturbation to the strain polarisations as $h^{\rm{stoch}}$, which, for the + polarization, reads
\begin{align}
    \label{eq:Harmonics}
    h^{\rm{stoch}}_{+}&=h_{0}\delta \sum^{\infty}_{n=1} \frac{(A_{n}^{r}+B_{n}^{\phi}) (2 l+n)^2 }{4 l^2}\cos \left( \frac{n \omega }{l}t+2 \omega t\right) \nonumber \\
    &+ h_{0}\delta \sum^{\infty}_{n=1} \frac{(A_{n}^{r}-B_{n}^{\phi}) (2 l-n)^2 }{4 l^2} \cos \left( \frac{n \omega }{l}t-2 \omega t\right) \nonumber \\
    &- h_{0}\delta \sum^{\infty}_{n=1} \frac{(A_{n}^{\phi}-B_{n}^{r}) (2 l+n)^2 }{4 l^2} \sin \left( \frac{n \omega }{l}t+2 \omega t\right) \nonumber \\
    &- h_{0}\delta \sum^{\infty}_{n=1} \frac{(A_{n}^{\phi}+B_{n}^{r}) (2 l-n)^2 }{4 l^2} \sin \left( \frac{n \omega }{l}t-2 \omega t\right) \nonumber \\
    &+ \mathcal{O}\left[\delta^2 \right],
\end{align}

\noindent whereas for the $\times$ polarisation the labels are switched ($r \to \phi$ and $\phi \to r$).

With the aid of Eq.~\eqref{eq:Harmonics}, alongside the relation between the torque fluctuations and the phase perturbations (Eqs~\ref{eq:coeffeq1} and \ref{eq:coeffeq2}), we can determine the correspondence between a torque fluctuation curve of a source embedded in an accretion disc and the harmonic content of its GWs. For the remainder of this paper, we will refer to this extra harmonic content as \textit{dirty waveforms} (DWs), as opposed to the ``clean'' waveform that one would expect from a vacuum GW source.

Recalling our scaling for the stochastic torque fluctuations (Eqs~\ref{eq:fluctscaling1} and \ref{eq:fluctscaling2}), we find that the size of a typical DW, at a frequency equal to $n/l$ times the orbital frequency, reads
\begin{align}
    \label{eq:DWampli}
   \lvert  h_{n}^{\rm{DW}}\rvert \sim \sigma_{n}\frac{4 G \dot{L}_{\rm{lin}}}{c^4 D_{\rm{L}}},
\end{align}

\noindent where we neglected an order-unity factor of $(2l \pm n)^2/n^2$ for simplicity. It is worth noting here that Eq.~\eqref{eq:DWampli} is a general result and can be used to model any kind of stochastic or unmodelled torque fluctuations, although some care is necessary to model the mass quadrupole properly.

We now show the magnitude of the typical DWs caused by gas torque fluctuations, scaled for our simple disc model (see Section~\ref{sec:Methods:disk}). For Type I torques, they read
\begin{align}
   \lvert h_0^{-1} h^{\rm{DW}}_{n} \rvert &\sim 2.4 \times 10^{-6} \sigma_{n} q  \left(\frac{M}{10^6 \, \rm{M}_{\odot}}\right) \nonumber\\ &\times \left(\frac{10^{-4}\,\rm{Hz}}{f_{\rm z}}\right)\left(\frac{\rho_0}{10^{-5} \rm{g}\,\rm{cm}^{-3}}\right)
\end{align}

Conversely, for Type II torques, they read
\begin{align}
  \lvert h_0^{-1} h^{\rm{DW}}_{n} \rvert &\sim 1.9 \times 10^{-10} \frac{\alpha \sigma_n}{q} \left(\frac{M}{10^6 \, \rm{M}_{\odot}}\right) \nonumber\\ &\times \left(\frac{10^{-4}\,\rm{Hz}}{f_{\rm z}}\right)\left(\frac{\rho_0}{10^{-5} \rm{g}\,\rm{cm}^{-3}}\right).
\end{align}

Looking at these two equations, we find that DWs can vary from a few times to several orders of magnitude smaller than the main carrier GW signal of the source. Not surprisingly, the relative amplitude of DWs grows linearly with the density normalisation of the disc. Furthermore, heavier mergers produce louder DWs relative to their main GW signal. Finally, lower-frequency sources also produce louder DWs. Note that the latter two scaling relations depend on the disc model assumptions. However, in the same vein as in Section~\ref{sec:Methods:disk}, we take them to be sufficient as an order-of-magnitude estimate.

While undoubtedly small in amplitude, DWs contain frequency information that can be both lower and, crucially, higher than the frequency of the carrier GW itself. This has the interesting implication that loud, low-frequency sources that would normally not fall within the LISA frequency band can potentially produce DWs that do. We discuss this in more detail in Section~\ref{sec:Detectability}, where we quantify the detectability of DWs both for individual sources and as a stochastic GW background.

\subsection{Secular energy flux}\label{sec:DirtyWaveforms:flux}

In the previous section, we found that torque fluctuations can be imprinted into some additional harmonic content in the GWs of a gas-embedded inspiral. Here we derive a new \textit{secular} effect, that corresponds to the extra energy being carried away by DWs.\footnote{Dirty emissions, if you will.} Luminosity is proportional to the square of a wave's amplitude and, by analogy, we expect the additional energy flux to be a quadratic effect in the perturbation.

We use the quadrupole formula \citep[][]{Einstein1916} and insert the perturbed quadrupole moment tensor
\begin{align}
    \left<\frac{{\rm d}E}{{\rm d}t}\right> &=\left<\frac{G}{5c^5 } \frac{{\rm d}^3\mathbf{Q}}{{\rm d}t^3}\frac{{\rm d}^3\mathbf{Q}}{{\rm d}t^3}\right> \\
    & \approx  \left<\dot{E}_{\rm q}\right> +  \delta^2\left<\dot{E}_{\rm DW}\right>,
\end{align}

\noindent where the linear order contribution vanishes because of the orbit average. After a lengthy but straightforward computation we find
\begin{align}
    \dot{E}&=  \dot{E}_{\rm{q}}\left(1 +\delta^2 \sum_{n=1}^{\infty} \mathcal{P}_{n}^{\rm{DW}}(A_{n}^{r},B_{n}^{r},A_{n}^{\phi},B_{n}^{\phi},n) \right) +\mathcal{O}[\delta^3],
\end{align}

\noindent where the $\mathcal{P}_{n}^{\rm{DW}}$ denote the contributions of all different DW harmonics to the secular energy flux of the binary source. They are given by
\begin{align}
    \mathcal{P}_{n}^{\rm{DW}}&= \frac{1}{32l^6} \left({A_{n}^{\phi}}^2 + {B_{n}^{\phi}}^2 \right)\left(240 n^2l^4 + 60 n^4l^2+ n^6\right)  \nonumber\\
    &+ \frac{1}{4l^6}\left(A_{n}^{r} B_{n}^{\phi}  - A_{n}^{\phi}  B_{n}^{r} \right)\left( 48nl^5 + 40 n^3l^3 + 3 n^5l\right)\nonumber\\
    &+\frac{1}{24l^6} \left({A_{n}^{r}}^2 + {B_{n}^{r}}^2 \right)\left(72l^6 + 180n^2l^4 + 45 n^4l^2 + n^6\right),
\end{align}

\noindent which reduces to a simpler form if we assume that the perturbations are sourced by torque fluctuations (Eqs~\ref{eq:coeffeq1} and \ref{eq:coeffeq2}):
\begin{align}
 \mathcal{P}_{n}^{\rm{DW}}&= \frac{\left({A_{n}^{\tau}}^2 + {B_{n}^{\tau}}^2 \right)}{32 n^4 a^4 \mu^2 \omega_{\rm o}^4 l^6} \left(240 n^2l^4 + 60 n^4l^2+ n^6\right)
\end{align}

Note that higher-frequency harmonics contribute strongly to the total emitted power due to the $\sim$$n^2$ scaling. This is expected, because the GW luminosity depends on the third derivative of the quadrupole moment, meaning that high-frequency motion is weighted very strongly. The total relative power emitted through DWs will be the sum over all harmonic contributions:
\begin{align}
    \mathcal{P}^{\rm{DW}} = \sum^{\infty}_{n=1} \mathcal{P}_{n}^{\rm{DW}}.
\end{align}

To reiterate: we find that stochastic torque fluctuations can cause an additional secular energy flux, caused by the additional harmonic content present in the GWs of the binary system. This is in addition to the energy loss caused by linear torques presented in Section~\ref{sec:Methods:Linear}. The effects persist independently at quadratic order, \textit{even if the net torque vanishes} when averaged over the appropriate amount of orbits.

To investigate the magnitude of the added energy flux, we insert our simple disc model (Eqs~\ref{eq:diskmodel} and \ref{eq:diskmodelcs}). For Type I and Type II torques, respectively, we find the scaling relations
\begin{align}
    \mathcal{P}^{\rm{DW}}_{n} &\approx 2.8 \times 10^{-12} p(n^2) \sigma_{n}^2 q^2 \left( \frac{10^{-4} \, \rm{Hz}}{f_{\rm{z}}} \right)^{2} \nonumber \\ 
    &\times \left(\frac{M}{10^6 \, \rm{M}_{\odot}} \right)^{2}\left( \frac{\rho_0}{10^{-5} \, \rm{g}\,\rm{cm}^{-3}} \right)^2
\end{align}

\noindent and
\begin{align}
    \mathcal{P}^{\rm{DW}}_{n} &\approx 9.2 \times 10^{-22} \alpha^2 p(n^2) \frac{\sigma_{n}^2}{q^2} \left( \frac{10^{-4} \, \rm{Hz}}{f_{\rm{z}}} \right)^{2} \nonumber \\ 
    &\times \left(\frac{M}{10^6 \, \rm{M}_{\odot}} \right)^{2}\left( \frac{\rho_0}{10^{-5} \, \rm{g}\,\rm{cm}^{-3}} \right)^2,
\end{align}

\noindent where the polynomial $p(n^2)$ is given by
\begin{align}
p(n^2) = \frac{1}{l^6 n^4} \left(240 n^2l^4 + 60 n^4l^2+ n^6\right).
\end{align}

As we mentioned before, an additional radiated power produces a dephasing in the waveform proportional to the observation time and the rest-frame frequency of the source (Eq.~\ref{eq:depheq}). Since every DW harmonic contributes to an amount of dephasing $ \delta \phi^{\rm{DW}}_{n}$, the total amount of dephasing is given by the sum
\begin{align}
    \delta \phi^{\rm{DW}}= \sum_{n=1}^{\infty} \delta \phi^{\rm{DW}}_{n} \approx \frac{1}{2} f_{\rm z} \mathcal{P}^{\rm{DW}} T_{\rm{obs}},
\end{align}

\noindent where we assume that the source is approximately monochromatic within one observation time. Using our disc model and the simple torque formulae, we find that for Type I torques the different dephasing contributions amount to
\begin{align}
    \label{eq:dephasingTI}
    \delta\phi^{\rm{DW}}_{n} &\approx 3.4 \times 10^{-10}p(n^2) \sigma_{n}^2 q^2 \left( \frac{10^{-4} \, \rm{Hz}}{f_{\rm{z}}} \right)^{1/3}\nonumber \\ 
    &\times \left(\frac{M}{10^6 \, \rm{M}_{\odot}} \right)^{8/3}\left( \frac{\rho_0}{10^{-5} \, \rm{g}\,\rm{cm}^{-3}} \right)^2 \left( \frac{T_{\rm{obs}}}{1 \, \rm{yr}} \right),
\end{align}

\noindent whereas for the Type II torque regime we find
\begin{align}
    \label{eq:dephasingTII}
     \delta \phi^{\rm{DW}}_{n} &\approx 9.6 \times 10^{-19} \alpha^2 p(n^2) \frac{\sigma_{n}^2}{q^2} \left( \frac{10^{-4} \, \rm{Hz}}{f_{\rm{z}}} \right)^{1/3} \nonumber \\ 
    &\times \left(\frac{M}{10^6 \, \rm{M}_{\odot}} \right)^{8/3}\left( \frac{\rho_0}{10^{-5} \, \rm{g}\,\rm{cm}^{-3}} \right)^2\left( \frac{T_{\rm{obs}}}{1 \, \rm{yr}} \right).
\end{align}

\begin{figure}
    \centering
    \includegraphics[scale=0.7]{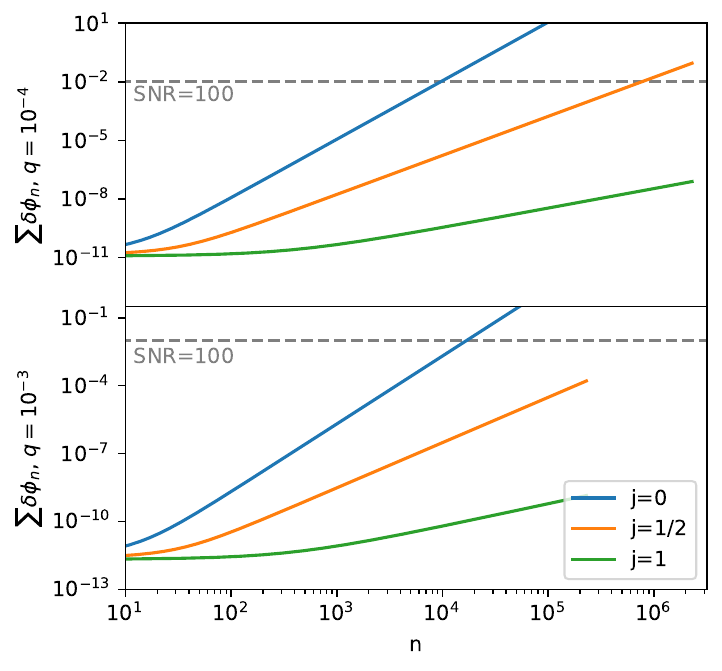}
    \caption{We show the value of the accumulated dephasing $\delta \phi$ that is caused by all DW harmonics up to $n$ times the orbital frequency, for different fluctuation power-law models (coloured lines). Here we scale the dephasing contributions (Eqs~\ref{eq:dephasingTI} and \ref{eq:dephasingTII}) with typical values of $M = 10^6$ M$_{\sun}$, $\rho_0=10^{-5}$~g~cm$^{-3}$, $\sigma = 10^2$, and one year of observation time. We see that, if stochastic torque fluctuations preserve their amplitude for high harmonic numbers, the dephasing caused by DWs could potentially fall within the detectable range for LISA. A dashed horizontal line denotes the dephasing that is likely detectable (see the text under Eq.~\ref{eq:dephasingTIIlin}) for a source with an SNR of 100. We discuss the detectability in more detail in Section~\ref{sec:Detectability:individual}.
    }
    \label{fig:accdephasing}
\end{figure}

Figure~\ref{fig:accdephasing} shows an example of how the total dephasing accumulates as more and more harmonics are considered, for different power-law scalings of the torque fluctuation amplitudes. The total dephasing caused by DWs can potentially reach fractional values for sources that cycle in the LISA band, depending on the value of the fluctuation spectral index $j$. If we allow for the central density to exceed our standard estimate of $\rho_0 \sim 10^{-5}$~g~cm$^{-3}$ (for a primary BH of $10^6$~M$_{\sun}$), significant dephasing is even more likely.

Considering that it is always the high-frequency fluctuations that dominate the total dephasing, we can estimate the sum of harmonics as follows:
\begin{align}
    \label{eq:sum}
    \sum_{n=0}^{n_{\rm{max}}} p(n^2) \sigma_n^2 \approx \sigma^2 \sum_{n=0}^{n_{\rm{max}}} \frac{n^2}{l^6} \left( \frac{l}{n} \right)^{2j} \sim \sigma^2 \frac{n_{\rm{max}}^{3-2j}}{3-2j} \frac{1}{l^{6-2j}},
\end{align}

\noindent where, in the last step, we assumed that $3/2<j \leq 0$ and expanded in large values of $n_{\rm{max}}$. If we insert Eq.~\eqref{eq:nmax} for $n_{\rm{max}}$ in Eq.~\eqref{eq:sum}, we find that the dephasing can indeed accumulate strongly when summed over all harmonics. The total dephasing is larger by $(3-2j)$ factors of $n_{\rm{max}}$ (which is of the order of $\sim 2.3 \times 10^2 q^{-1})$ than a single contribution. This amounts to a number that is of the order of $10^{9-18}$ for $j=0$, $10^{6-12}$ for $j=1/2$ and $10^{3-6}$ for $j=1$, depending on the exact mass ratio. 

\section{Prospects for MILLI-Hertz Gravitational wave Detectors}\label{sec:Detectability}

In this section, we discuss the relevance and detectability of DWs in the context of GW detectors, in particular LISA. In Section~\ref{sec:Detectability:individual}, we show that the DW characteristic strain of very heavy mergers can shine in the LISA band, and that the accumulated dephasing caused by DW harmonics can be resolvable by LISA under favourable conditions. In Section~\ref{sec:Detectability:background}, we show that gas-embedded binaries produce a stochastic DW background that is comparable to other noise contributions in LISA's sensitivity curve, and that can reach detectable SNR values in four years of observation time. We find that all of these prospects depend strongly on the spectral index $j$, and become relevant for LISA whenever $j \lesssim 1/2$.

\subsection{Individual sources}\label{sec:Detectability:individual}

\subsubsection{Loud, low-frequency mergers}

The first question that we wish to answer is whether it would be possible to detect DWs directly within the data stream of LISA. The simplest measure of the SNR of a GW source is the characteristic strain, $h_{\rm{c}}$, and its height above the LISA strain sensitivity curve. It counts the number of GW cycles that can be filtered out from the noise and for a source in vacuum it is given by the following equation \citep[see, e.g.][]{Robson2019}:
\begin{align}
    h_{\rm c} = \sqrt{f_{\rm{z}}T_{\rm{obs}}}h_0 = \sqrt{f_{\rm{z}}T_{\rm{obs}}} \frac{f_{\rm{z}}^{2/3} (G M)^{5/3} \pi ^{2/3} q}{c^4 D_{\rm{L}}},
    \label{eq:hc}
\end{align}

\noindent where we expanded in small mass ratios and dropped a prefactor of order of a few. Here we assumed that the source is approximately monochromatic for the duration of one observation time, and we can therefore extract the full amount of cycles, $\sqrt{f_{\rm{z}}T_{\rm{obs}}}$.

For a gas-embedded source that produces DWs, we can build a spectrum of strains, that correspond to the different harmonics present in the GW signal. Recalling the typical size of the physical DWs (Eq.~\ref{eq:DWampli}), we find the following formula for the characteristic strain of a DW harmonic with a frequency of $f_{\rm z} n/l$:
\begin{align}
    \label{eq:hcdirty}
    h_{{\rm c}, n}^{\rm{DW}}(f_{\rm{z}}n/l)= \left( f_{\rm{z}}T_{\rm{obs}} \frac{n}{l}\right)^{1/2} \sigma_{n}\frac{G \dot{L}_{\rm{lin}}}{c^4 D_{\rm{L}}}.
\end{align}

Here we imply that the amplitude of every DW harmonic can be considered constant for the length of one observation time, an assumption that requires some further discussion. Stochastic torque fluctuations are sourced by the physics of the gas surrounding the GW source. For this reason, we should expect their amplitude to vary on time-scales that are completely separate from the gravitational radiation time-scale or the duration of LISA's observation run. However, as we have mentioned in Section~\ref{sec:Methods:Stochastic}, we are interpreting the amplitudes of the fluctuations as the typical pick from a distribution with a given variance $\sigma_n$. These amplitudes are selected every $l$ orbits, and will generally be of the order of the distribution's standard deviation. Therefore, Eq.~\eqref{eq:hcdirty} can also be interpreted in a similar fashion: it denotes the characteristic strain corresponding to a statistically typical realisation of DWs, for a GW source embedded in a thin disc. Once again, we believe that this idealisation suffices as an order-of-magnitude estimate and more precise characterisations of the DW strain spectrum can be constructed as soon as the fluctuation amplitudes are better constrained by hydrodynamical simulations.

A quick glance at Eq.~\eqref{eq:DWampli} shows that typical LISA sources (BH binaries with a total mass of $10^5$--$10^7$~M$_{\sun}$) produce DWs that are far too small to be detected by planned GW observatories. However, it also shows that the DW amplitude strongly scales with the total mass of the system. BH mergers with a total mass exceeding $10^7$ M$_{\sun}$ produce loud GWs, but merge at frequencies lower than what LISA is able to detect. DWs are, however, not limited to the frequency of an orbit around the primary BH. Rather, they can persist all the way up to frequencies corresponding to an orbit around the ISCO of the secondary (see Eq.~\ref{eq:nmax} and the discussion around it). Low-frequency GW sources can therefore produce DWs in a more appropriate frequency band for LISA.

\begin{figure}
    \centering
    \includegraphics[scale=0.7]{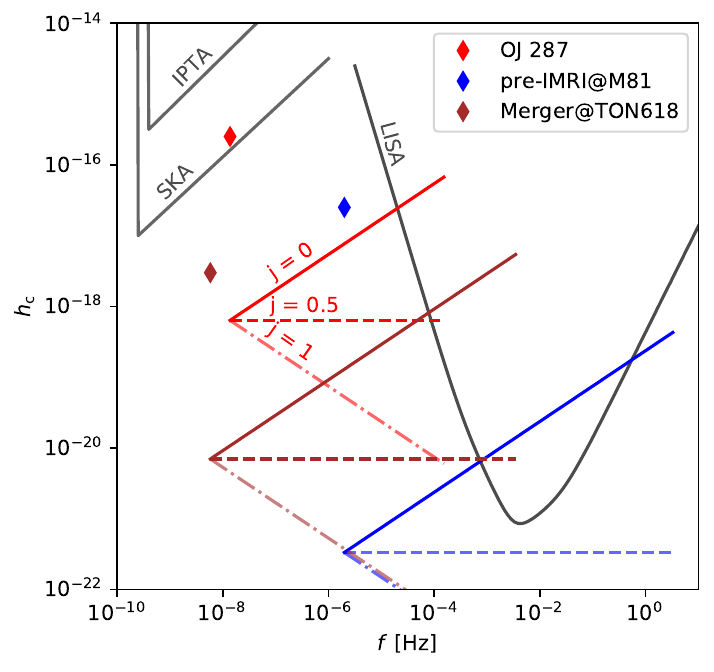}
    \caption{We show the characteristic strain of some GW sources embedded in the accretion discs of observed quasars/AGN (coloured diamonds) for one year of observation time. For the quasar OJ 287, we choose the best fit model for the secondary SMBH \citep[][]{Laine2020}, which yields a mass ratio of $\sim 10^{-3}$. For the other two sources, we imagine a hypothetical secondary companion with a mass ratio of $\sim 10^{-4}$. We plot our estimates for the DW characteristic strain spectra. If the spectral index $j$ is of the order of $\lesssim 1/2$, we find that our hypothetical (but realistic) sources can produce DWs that are loud enough to shine in the LISA band. Especially interesting is OJ 287, an SMBH binary candidate which could be loud enough to be a resolvable single source for future PTA searches. If its DWs were also detectable, OJ 287 could be the first nHz-mHz multiband source of GW waves.}
    \label{fig:hcbigbois}
\end{figure}

In Figure~\ref{fig:hcbigbois}, we show some realistic examples where the main GW signal of an SMBH binary is at nHz frequencies, but its DWs can shine all the way up to the mHz band. If the fluctuation amplitudes are large enough at higher harmonic numbers, it is possible for DWs to reach LISA's level of strain.

An interesting case study is the quasar OJ 287, which perfectly illustrates the type of source that could produce audible DWs.
OJ 287 likely consists of an SMBH binary with total mass of $\sim 2\times 10^{10}$ M$_{\sun}$ and mass ratio of $\sim 10^{-3}$, and the periodic variations in its luminosity can be explained by the secondary's interaction with the accretion disc \citep[][]{Dey2018,Laine2020}. This system is particularly exciting because it could, due to its large strain and convenient frequency, be one of the prime targets for a single source detection with future pulsar timing array (PTA) observations with the square kilometer array \citep[see, e.g.][]{Moore2015,Zhu2015}. Interestingly, its DWs cross over LISA's sensitivity curve for a spectral power index of $j \lesssim 1/2$. If this is the case in reality, OJ 287 could be a candidate for the first nHz-mHz multiband source of GW waves. Another observed binary candidate with very similar characteristics is the quasar PSO J334.2028+01.4075 (\citealt{Liu2015}, although see \citealt{Benke2019}) at redshift $z\sim 2$, among several candidates found by quasar periodicity searches \citep[][]{Charisi2016,Liu2019,ChenY2020}. We discuss the intriguing observational prospects of such PTA-LISA multiband detections in Section~\ref{sec:Discussion}. The other two examples shown in the plot consist of a hypothetical SMBH in orbit around the quasar TON618, and a hypothetical IMBH orbiting around M81. In both of these examples, the main GW signal would be inaudible to LISA (or PTA), while the DW characteristic strain can reach LISA's sensitivity.

We stress that we are not claiming that these particular GW sources must, or are even likely to produce audible DWs. Rather, we only show that it is indeed possible for the DWs of a \textit{realistic} heavy binary to shine through LISA's sensitivity band. Computing whether LISA will actually be likely to detect similar systems requires a more thorough analysis of event rates and SNRs. Furthermore, it will also have to wait for more precise constraints on the fluctuation spectra. Nevertheless, the possibility of detecting such systems with LISA is very intriguing (see Section~\ref{sec:Discussion} for further discussion), and deserves further development.

\subsubsection{Dephasing of LISA sources}

In this section, we investigate whether the effects of DWs are observable in  LISA's more conventional targets, which are binaries with a total mass of approximately $10^5$--$10^7$~M$_{\sun}$. While the DW strain is too small for direct detection, we have shown in Section~\ref{sec:DirtyWaveforms:flux} how the total dephasing caused by DWs can accumulate to a substantial amount when the contributions from all harmonics are considered. 

\begin{figure}
    \centering
    \includegraphics[scale=0.55]{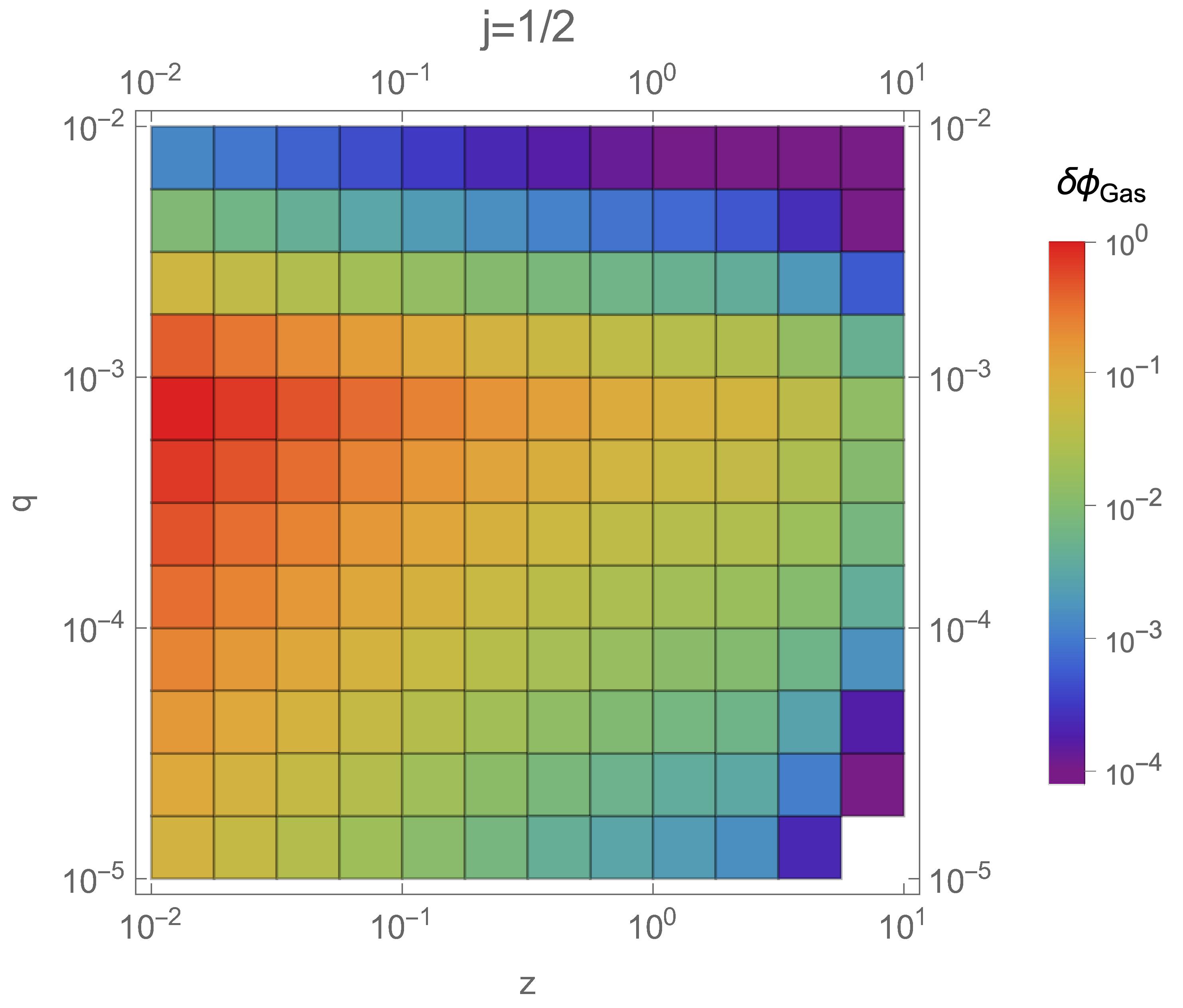}
    \caption{We show the accumulated dephasing caused by DWs of a standard LISA source with total mass of $10^6$ M$_{\sun}$ embedded in the simplified disc model described by Eqs~\eqref{eq:diskmodel} and \eqref{eq:diskmodelcs}, for a wide range of mass ratios and redshifts. The dephasing is calculated from the amount of time for which the characteristic strain (Eq.~\ref{eq:hc}) of the source remains above the LISA sensitivity curve, along with Eqs~\eqref{eq:dephasingTI} and \eqref{eq:dephasingTII}. For this plot, we focus on the spectral index $j=1/2$. Clearly visible is the transition area between the Type I and Type II torque regimes close to a mass ratio of $q\sim 7\times 10^{-4}$, for which the dephasing is significant for a large range of redshifts. Note that this effect is in addition to the dephasing caused by linear torques (Eqs~\ref{eq:dephasingTIlin} and \ref{eq:dephasingTIIlin}), and can be present even if the average of the torque fluctuations vanishes. The shape of the contours of the plot is unaffected by the choice of $j$, while the normalisation of the colour bar changes by orders of magnitudes (see text). The white square is a special case in which the strain of the source never reaches LISA's sensitivity.}
    \label{fig:dephj}
\end{figure}

In Figure~\ref{fig:dephj}, we show the accumulated dephasing due to the DW harmonics that is expected for a typical LISA source, as a function of its redshift and mass ratio. To represent the typical source, we chose a total mass of $10^6$ M$_{\sun}$ embedded in the simplified disc model presented in Section~\ref{sec:Methods:disk}, with a viscosity parameter $\alpha =0.01$. To estimate the dephasing, we use the standard procedure of calculating the amount of time for which the characteristic strain (Eq.~\ref{eq:hc}) of the source remains above the LISA sensitivity curve. This yields a total observation time, capped at a LISA observation window of four years, which we can insert into the dephasing Eqs~\eqref{eq:dephasingTI} or \eqref{eq:dephasingTII}. We focus on a fluctuation power-law model with $j=1/2$, since it produces some interesting, marginally-detectable cases. Furthermore, the shape of the contours in the plot are preserved for any value of $j$, total mass, and disc density normalisation.

As we can see in Figure~\ref{fig:dephj}, higher-redshift sources dephase less, simply because they spend less time in the LISA band at low frequencies. Maximal dephasing is achieved in the vicinity of the transition mass ratio (Eq.~\ref{eq:qcrit}) between Type I and Type II torques. For the case of $j=1/2$, an appreciable dephasing is reached for a large range of redshifts when the mass ratio is between $\sim 10^{-5}$ and $\sim 10^{-3}$. Setting $j=0$ increases the dephasing by a factor $\sim 10^5$ for the entire parameter space we are considering. Note that such very large values for the dephasing are better understood as a change in the inspiral time-scale. Setting $j=1$ decreases it by a similar amount. Increasing the total mass of the primary or the disc density parameter also increases or decreases the dephasing according to Eqs~\eqref{eq:dephasingTI} and \eqref{eq:dephasingTII}. Once again, these results suggest that the accumulated dephasing caused by DW harmonics will be of sufficient magnitude to appear along the linear torque effect, provided that torque fluctuations preserve their amplitude beyond the resolution limit of current hydrodynamical simulations.

\subsection{Gravitational wave background}\label{sec:Detectability:background}

As we have seen, low-frequency GW sources can produce higher-frequency DWs if they happen to be buffeted by stochastic torque fluctuations. While DWs are small in amplitude, the can shine in the mHz band, where LISA is the most sensitive. Note that, given a total mass, there should be many more low-frequency binaries than high-frequency ones, simply because the residence time at a given separation scales strongly in the case of GW emission \citep[][]{Peters1963}. Although this simple picture can be affected by many environmental effects \citep[see, e.g.][for a recent analysis]{Bortolas2021gas}, it leads us to suspect that the sum of the DWs produced by all low-frequency GW sources that happen to be embedded in an accretion disc might produce a stochastic GW background relevant for LISA.

To estimate the total DW stochastic background, we follow the prescription detailed in \cite{Bonetti2020} \citep[based on the works of][]{Finn2000,Phinney2001}, which is applied to the higher harmonics of eccentric extreme mass-ratio inspirals. In the case of DWs, the calculations are closely analogous, but the source of the harmonics are torque fluctuations rather than orbital eccentricity.
The total GW background $h_{\rm{bg}}^{\rm{DW}}$ adds in quadrature by summing over all gas-embedded sources that produced a DW harmonic which reaches Earth with a frequency $f$:
\begin{align}
    \label{eq:BGintegral}
     h^{\rm{DW}}_{\rm{bg}}(f)^2&= \int \,{\rm d}z\, {\rm d}M\, {\rm d}q  \nonumber \\ &\times \sum_{n=1}^{n_{\rm{max}}} \left[\frac{{\rm d}^4 N_{\rm{GE}}}{{\rm d}\log(f_{\rm{s}}) \, {\rm d}z \, {\rm d}M \, {\rm d}q}  h^{\rm{DW}}_{n}(f)^2\right]_{f_{\rm s}=f_{\rm z}/n},
\end{align}

\noindent where $N_{\rm{GE}}$ denotes the total number of gas-embedded merger events in the Universe. To make progress, we separate out the differential contributions to the total number of gas-embedded mergers as follows \citep[see, e.g][]{wetEMRI2021}:
\begin{align}
    \frac{{\rm d}^4 N_{\rm{GE}}}{{\rm d}\log(f_{\rm{s}}) \, {\rm d}z \, {\rm d}M \, {\rm d}q} =  \frac{1}{\dot{f}_{\rm s}} \frac{1}{1+z} \frac{{\rm d}N_{\rm{AGN}}}{{\rm d}M}\frac{{\rm d}\dot{\Gamma}}{{\rm d}q}\frac{{\rm d} V_{\rm z}}{{\rm d}z}, 
\end{align}

\noindent where $N_{\rm{AGN}}$ is the total number of primary BHs that host an accretion disc, $\dot{\Gamma}$ is the merger event rate per SMBH, $V_{\rm z}$ is the cosmological volume element, and the factor $1/\dot{f}_{\rm s}$ accounts for the residence time of the binary at a particular frequency.

From this point on, there are many possibilities for the different ingredients that compose the differential number density. We will make some plausible assumptions, and will dedicate several paragraphs within the discussion (Section~\ref{sec:Discussion}) to explore the caveats and the limitations of our choice. We believe that our estimates represent a reasonable order-o-magnitude estimation of the importance of DWs as a GW background, but more work has to be done to constrain it precisely.

First, we assume that the mass function of SMBHs that host an accretion disc can be simply described by a factor $r_{\rm{AGN}}$ multiplied by the SMBH mass function:
\begin{align}
\label{eq:ragn}
 \frac{{\rm d}N_{\rm{AGN}}}{{\rm d}M} = r_{\rm{AGN}}  \frac{{\rm d}N_{\bullet}}{{\rm d}M}.
\end{align}

The value of $r_{\rm{AGN}}$ is estimated to be of order 0.01 in the local Universe, and of order 0.1 at $z \sim 2$ \citep[][]{Greene2007,Macuga2019}. As an estimate for the mass function, we use a simple phenomenological fit of the data in \citet{Greene2007} that is commonly used for LISA estimates \citep[see, e.g.][]{Gair2010}:
\begin{align}
    \frac{{\rm d}N_{\bullet}}{{\rm d}M}=  \frac{0.0018 \left( \frac{M}{\rm{M}_{\odot}} \right)^{0.03} }{1+7.9\times 10^{-12} \left(\frac{M}{\rm{M}_{\odot}}\right)^{3/2}} \, \left[\rm{Mpc}^{-3} \right]
\end{align}

Secondly, we need to choose a formation channel for gas-embedded mergers. From Section~\ref{sec:DirtyWaveforms:harmonics}, we know that the loudest DWs are produced by high-mass merger events, which are likely the result of galaxy mergers \citep[see, e.g.][]{Begelman1980,Hopkins2006,Mayer2007,Capelo2015}. We use the halo merger rate from the Millennium simulations \citep[][]{Fakhouri2010} as a tracer for galaxy mergers, which in turn should trace the mergers of heavy BH binaries. However, not all halos host an SMBH. Similarly to Eq.~\eqref{eq:ragn}, we assume that the SMBH merger rate can be expressed by multiplying the halo merger rate by a simple factor, the occupation number $r_{\rm{occ}}$, which is estimated to be of order $\gsim 0.1$ \citep{Lippai2009}. Moreover, we introduce a factor $r_{\rm{al}}$ which describes the amount of binaries in which the secondary will have aligned itself with the disc by the time it reaches the GW-dominated regime. \citet{Fabj2020} show that, for realistic AGN models, stellar-mass compact objects can be dragged into alignment with the disc efficiently if their inclination is below $\sim 15^{\circ}$ to $\sim 30^{\circ}$. Here we assume that this result can be extended to a larger range of small mass-ratio binaries. In the case of equal-mass SMBH mergers, the circumbinary disc and the orbital plane similarly align themselves on a time-scale of roughly 10$^6$~yr if they are misaligned at separations of $\sim 10^4$ $r_{\rm S}$, and faster by a factor $\sim q^{-1}$ for smaller mass ratios \citep[][]{Larwood1997,Miller2014}. Given these considerations, we estimate that the alignment factor is roughly of fractional order for any mass ratio, although we expect a more precise estimate to depend on the binary and disc properties. The merger event rate is finally given by
\begin{align}
    \frac{{\rm d}\dot{\Gamma}}{{\rm d}q} = r_{\rm{occ}} r_{\rm{al}} \frac{{\rm d}\dot{\Gamma}_{\rm{Halo}}}{{\rm d}q}.
\end{align}

For the halo merger rate, we adopt the simple fit found in \citet{Fakhouri2010}:
\begin{align}
    \label{eq:mergerrate}
    \frac{{\rm d}\dot{N}_{\rm{Halo}}}{{\rm d}q} &= \dot{z}\frac{{\rm d}^2N_{\rm{Halo}}}{{\rm d}q{\rm d}z}\\
    &\approx \dot{z}\frac{0.01}{q^2} \left(\frac{M_{\rm{Halo}}}{10^{12} \,\rm{M}_{\odot}} \right)^{0.13}\exp\left( \frac{q}{10^{-2}}\right)^{0.26}(1+z)^{0.1},
\end{align}

\noindent where the derivative of the redshift parameter reads \citep[][]{Peebles1993}
\begin{align}
    \label{eq:cosmology}
    \dot{z}= -(1+z)H(z) = 4 \pi D_{\rm{c}}^2(1+z)^{-2} c \left(\frac{{\rm d}V_{z}}{{\rm d}z} \right)^{-1}.
\end{align}

Here $H(z)$ is the Hubble parameter and $D_{\rm c}$ is the comoving distance. Using the latter two equations simplifies the differential merger rate expression to
\begin{align}
    \frac{{\rm d}^4 N_{\rm{GE}}}{{\rm d} \log(f_{\rm{s}})\, {\rm d}z \, {\rm d}M \, {\rm d}q} =\mathcal{R} \frac{c}{\dot{f}_{\rm s}} \frac{4 \pi D_{\rm{c}}^2}{(1+z)^3} \frac{{\rm d}N_{\bullet}}{{\rm d}M}\frac{{\rm d}^2N_{\rm{Halo}}}{{\rm d}q {\rm d}z},
\end{align}

\noindent where we grouped all unconstrained fractional order prefactors within a single factor $\mathcal{R}= r_{\rm{AGN}}r_{\rm{occ}}r_{\rm{al}}$. Now we have to relate the mass of the halos that are merging to the mass of the SMBH that they are hosting. For this purpose, we use a simplified power-law fit adapted from \citet{Croton2009}:
\begin{align}
   M_{\rm{Halo}} \sim \frac{2\times 10^7}{1+z} \left(\frac{M}{\rm{M}_{\odot}} \right)^{2/3}\rm{M}_{\odot}.
\end{align}

Finally, we assume that secondaries which are aligned with the disc of the primary do indeed decay according to Eq.~\eqref{eq:adot}, following the time-scales shown in Figure~\ref{fig:timescales}. Then we can express the frequency evolution as follows:
\begin{align}
    \dot{f} = \dot{f}_{\rm{gas}} + \dot{f}_{\rm{GW}} = \frac{3}{2}\sqrt{\frac{G M}{a^5}}\left(\dot{a}_{\rm{gas}}+\dot{a}_{\rm{GW}} \right).
\end{align}

Taking these assumptions at face value, we can evaluate the integral shown in Eq.~\eqref{eq:BGintegral} numerically by assuming our simplified disc model, the density parameter given by Eq.~\eqref{eq:diskdensity}, a viscosity parameter $\alpha=0.01$, and taking care to switch to the appropriate torque regimes for different mass ratios. 
The only integration limit that requires some care is for the mass ratio, where we integrate from the smallest value corresponding to a stellar-mass BH to a maximum of $1/25$, so that our torque models are appropriate. As we will discuss in Section~\ref{sec:Discussion}, mass ratios closer to unity could still contribute to the DW background due to the likely stochastic hydrodynamics in circumbinary discs.

\begin{figure}
    \centering
    \includegraphics[scale=0.7]{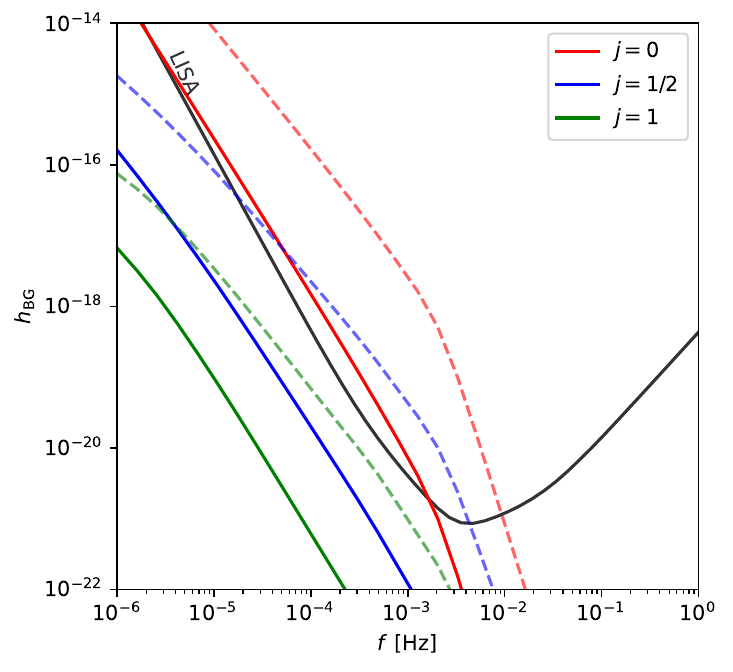}
    \caption{We show the DW background strain resulting from the integration of Eq.~\eqref{eq:BGintegral} for different fluctuation spectral indices. The low-frequency contribution is the result of DWs originating from the few but loud, heavy gas-embedded binaries ($M \gsim 10^9$ M$_{\sun}$) of the type discussed in Section~\ref{sec:Detectability:individual}. The high frequency background is produced by the more numerous lighter binaries ($M \lsim 10^8$ M$_{\sun}$). While the solid lines represent the instantaneous background, the dashed ones represent the amplitude of the background after four years of coherent integration, which reach an SNR of $\sim 10^3$, $\sim 20$ and $\sim 0.5$ for $j=0, j=1/2$ and $j=1$ respectively.}
    \label{fig:BG}
\end{figure}

Figure~\ref{fig:BG} shows the results for the expected DW background for different fluctuation power-law models, scaled for a reasonable choice of the factor $\mathcal{R}\sim 0.1 \times 0.1 \times 0.1 = 10^{-3}$. The main contribution to the DW background are very heavy gas-embedded binaries with a total mass of $\gsim 10^9$ M$_{\sun}$,  orbiting at frequencies of $\sim 10^{-8}$ Hz at moderate redshift ($z \sim 2$). These are very similar to observed systems such as OJ 287 and PSO J334.2028+01.4075, which we discussed in the previous section. Lighter sources contribute to the higher-frequency end of the background, producing a characteristic inflection point visible in the plot. Not shown is the behaviour at lower frequencies, where the DW background decreases as fewer and fewer binaries can be considered to be in the GW-dominated regime.

For a fluctuation spectral index of order $j \lsim 1$, the background could potentially be detectable in the mHz range in a LISA mission lifetime.
Following \cite{Moore2015,Bonetti2020}, we estimate the SNR of the background with the following formula:
\begin{align}
    {\rm SNR} ^2 = \int \frac{ h_{ {\rm BG,DW} }^2 }{f^2S_{{\rm n}}} {\rm d}f,
\end{align}

\noindent where $S_{\rm{n}}$ is the square noise power spectral density for LISA. We obtain SNR values of the order of $\sim 10^3$, $\sim 20$, and $\sim 0.5$ for $j=0, 1/2$, and $1$, respectively, for four years of integration time.

Due to the simplicity of our assumptions, both in the modelling of the fluctuation as well as the merger rate, it is hard to conclusively assess whether the DW background will be a significant source of noise for LISA, or if it can be realistically detected within one mission lifetime. However, our calculations suggest that a background of DWs must exist, and that its size could be comparable to the galactic binary population noise  \citep[][]{Bender1998,LISA2007,LISAwhite2020} or other unresolved sources \citep[see, e.g.][]{Enoki2007,Ashley2010,Bonetti2020}. The detection of such a background and the study of its amplitude, slope, and inflection points could shed some light on the physics of accretion discs, and we will spend several paragraphs discussing this in Section~\ref{sec:Discussion}.

\section{Discussion}\label{sec:Discussion}

Over the course of this paper, we have made several simplifying assumptions, some of which were arbitrary. Therefore, we wish to dedicate Section~\ref{sec:Discussion:caveats} to re-examining our choices and discussing the possible consequences of changing them. The main results of this work (not regarding the formalism itself) are the three possible observable signatures of DWs that could be detected by planned GW observatories in the near future. We take some time to discuss how they could be used to constrain disc parameters and population models in Section~\ref{sec:Discussion:observation}, along with some more general challenges of detecting signatures of the environment with GW detectors. Finally, in Section~\ref{sec:Discussion:other} we briefly touch on the significance of DWs for more speculative future GW detectors.

\subsection{Model assumptions and caveats}\label{sec:Discussion:caveats}

\subsubsection{Disc model}

In order to produce some quantitative results, we have parametrized a generic thin disc model with Eqs~\eqref{eq:diskmodel} and \eqref{eq:diskmodelcs}. We based our choice on the assumption that most realistic accretion discs can be described with an $\alpha$ viscosity prescription. In the context of this paper, this assumption is somewhat arbitrary, since the analytical estimates for Type I and Type II torques (Eqs~\ref{eq:typeI} and \ref{eq:typeII}) only depend on the value of the \textit{local} density. Arguably, we could have scaled our results with the density at the separation of the secondary rather than with a central value, and left the question of the disc model completely open.

A large variety in density and temperature profiles exists in the literature of accretion disc models. We chose a standard disc model, but acknowledge that discs may be more complex and diverse across AGN. 
As an example, $\beta$-disc models predict much higher surface densities at the same separation \citep[see, e.g.][]{LyndenBell1974,Hure2001,Derdzinski2021}, suggesting that global torques could be much stronger than what we have assumed in this work. Moreover, the recent analysis of M87's SMBH accretion disc by the Event Horizon Telescope Collaboration suggests that it is magnetically arrested, \citep[MAD disc; see, e.g.][]{MAD1974,MAD2003,EHT2021}. 
It is difficult to speculate how torques and their fluctuations depend on the choice of disc model. In general, we expect the strongest variability in colder and denser discs (i.e. models that describe bright quasars, such as \citealt{SirkoGoodman2003}).
Once the question of the torque fluctuation spectra is better constrained, it will be straightforward to repeat our calculations for more exotic disc models. If some of them can produce large torque fluctuations of the type we have discussed throughout this paper, it is possible that their DWs would constitute a ``smoking gun'' observational signature, useful to determine what type of disc is emitting them.

\subsubsection{Torque fluctuation model}

In order to make a connection with existing simulations, we normalised our model for the unresolved sub-orbital fluctuations (Eq.~\ref{eq:flucamplitude}) with their amplitude at the orbital frequency (denoted with $\sigma$). We then parametrized the amplitude of higher-frequency fluctuations as a power law with a spectral index $j$. In later sections, we have shown that, for $\sigma \sim 100$, most of the observational prospects of DWs become borderline detectable for LISA whenever $j \lesssim 1/2$. Here we expand on the interpretation of $\sigma$ and $j$, in anticipation that future simulations will constrain these fluctuations.

The parameters $j$ and $\sigma$ served as a convenient description over a wide frequency band, but in reality the fluctuations may have a more nuanced high-frequency dependence. For example, it is possible that the fluctuations occur at a few characteristic frequencies rather than with a continuous spectrum, similar to the binary-SMBH hypotheses that explain periodic quasar emission at multiples of the orbital period (e.g. \citealt{DOrazio2015,DOrazio2015Natur}). The critical aspect for the observational signatures of DWs is the amplitude of the highest-frequency harmonics: the additional dephasing shown in Section~\ref{sec:Detectability:individual} is completely dominated by the high frequency contributions. Similarly, the detectability of DWs in LISA, both from single sources and as a background, also depends on the amplitude fluctuations at frequencies many orders of magnitude higher than the carrier nHz GWs of very massive binaries. Future hydrodynamical simulations can shed light on super-orbital fluctuations at higher frequencies, constrain $j$, or uncover more stochastic behaviour. 
In any case, the choice of parametrization is mostly a matter of convenience. In actuality, the requirement for DWs to be observable in LISA can be stated as follows:

\begin{itemize}
    \item Suppose the existence of a physical process that can produce torque fluctuations with frequencies of the order of an orbit around the ISCO of a small secondary BH, with amplitudes of the order $10^{-1}$ to $10^{-2}$ of the estimates given by the linear torque formulae. Then this process would produce DWs, the effects of which would (in principle) be visible in the LISA band.
\end{itemize}

This requirement on the amplitude of very high-frequency harmonics has the consequence that DWs can only be completely constrained (or even ruled out) with extremely high-resolution 3D simulations. These will have to properly model all physical processes which might lead to fluctuations, such as gas hydrodynamics, magnetic effects, accretion and feedback, back reaction, relativistic effects and so on. However, considering that the challenge is computational rather than conceptual, we believe that it should be achievable in the near future, and that the prospects of this endeavour justify the effort.

\subsubsection{Binary parameters and orbital eccentricity}

For this work, we have assumed that the binary can circularise efficiently as soon as it enters the GW-dominated regime. While this is a reasonable assumption, we have noted in Section~\ref{sec:DirtyWaveforms:buffet} that recent literature suggests that some measure of eccentricity can be retained all the way down to the LISA band in gas-embedded sources, although numerical simulations have yet to confirm this for intermediate and extreme mass ratios. 

On top of their different vacuum GW signature, eccentric binaries can excite additional modes in the disc \citep[][]{Goldreich1980,Papaloizou2000,Cresswell2007}. The new modes cause additional torque contributions to appear, which in turn will alter $\dot{e}$ and $\dot{a}$. The interplay between GW-driven emission and the torques caused by the reaction of the gas to an eccentric perturber can possibly produce additional gas-driven signatures, which might be detectable if the source can complete enough cycles in the LISA band. While not focused on in this work, we believe that this area of research will produce many interesting consequences for LISA and other GW observatories.

\subsubsection{SMBH merger model}

In our calculation of the DW stochastic background, we have assumed that the merger rate of dark matter haloes can be used as a tracer for the rate of SMBH mergers. We have parametrized our lack of knowledge of the occupation fraction, AGN fraction, and alignment fraction in a single prefactor $\mathcal{R}$, for which we can only estimate the order of magnitude. Another possibility would be to follow \cite{Haiman2009} and more recently \cite{Soyuer2021}, where a merger  rate model is based on the observationally determined quasar luminosity function. While the latter approach has the advantage of being tied to observational results, there are also many unknown factors that must be accounted for, such as the fraction of unobserved AGN and the amount of mergers per AGN lifetime. In our model we have also assumed that all AGN discs can be approximately described by our simplified disc model, which is chosen to reproduce the density and temperature of an $\alpha$-disc. As we have noted before, it might be the case that DW emission is dominated by more rare instances of accretion discs that happen to be denser and colder than the typical case. Computing a more precise GW background that includes this eventuality would require knowledge of the distribution of different types of accretion discs, a calculation that is beyond the scope of this paper. Finally, we have neglected the existence of a delay between the halo merger and the eventual merger of the SMBH binary. This delay is known to modify the merger rates for LISA binaries, and becomes a larger effect at higher redshifts and lower BH masses \citep[e.g.][]{SouzaLima2017,Khan2018}. However, most of the DW background is produced by relatively low-redshift, very massive mergers, where the delay is expected to be smaller \citep[see, e.g.][]{Khan2016}. Therefore, we can claim ``a posteriori'' that we do not expect delays to make a substantial difference in the results.

While there are many ways to produce a more precise population model, we believe that our choices lead to a reasonably solid estimate of the magnitude of the GW background produced by DWs, at least in the context of this paper. The larger uncertainty in the final result is rather due to the unresolved torque fluctuation power spectrum. Once torque variability is better constrained, it will be straightforward to repeat our background calculations with more precise estimates for the SMBH merger rates.

\subsection{Observational opportunities}\label{sec:Discussion:observation}

An important caveat for any claim that an environmental effect might be discernible with GW detectors is the problem of degeneracies in parameter estimation. The aim of this section is, however, not to discuss this limitation in detail, but rather to take an exploratory stance on what types of measurements are made possible by the existence of DWs assuming that they are actually loud enough to be detected (at least in  principle). We refer the reader to some of the several insightful works that discuss the challenges of identifying environmental effects \citep[][]{Kocsis2011,Barausse2014,Cardoso2020}, and note that all the usual caveats apply to our discussion. 

\subsubsection{Additional dephasing}

As mentioned in Section~\ref{sec:Methods:Linear}, the dephasing of an inspiralling gas-embedded GW source can be used as a probe of the disc properties. However, it is possible that any environmental effect that influences the binary evolution might instead be misinterpreted as a source with different parameters (e.g. chirp mass). The standard way to disentangle environmental effects from system parameters is to observe the rate of dephasing accumulation as the source chirps through different frequencies. To break these degeneracies, we must understand the dependency of torques and torque fluctuations on the disc and binary parameters, i.e. we must quantify $\sigma = \sigma(M,q,\rho_0,f,\alpha, \, ...)$ and $j= j(M,q,\rho_0,f,\alpha, \, ...)$. This is an additional challenge to quantifying the diversity of average torque effects, which are also sensitive to disc parameters \citep{Derdzinski2021}. Nevertheless, the combined detection of a dephasing caused by linear torques alongside the dephasing caused by DWs could provide a stronger probe of accretion disc models than what is currently considered possible with GW detectors. This analysis will be easiest for the loudest, chirping sources and might not be possible for all detections.

\subsubsection{Stochastic background}

Of all the observational prospects, measuring the stochastic background caused by DW seems the most realistic. As for any GW background, its SNR can be accumulated over the whole duration of LISA's observing run without requiring prior knowledge of the constituent waveforms. If a DW background is observed with enough SNR, its amplitude, slope, and inflection points will contain a lot of (degenerate) information regarding torque fluctuations, disc properties, SMBH migration, and merger rates.
Here we explore two scenarios in which a detection of a DW background could yield interesting results.

In the first scenario, we imagine that, by the 2030s,  LISA and PTA observations will have determined the SMBH merger rate,
and that constraints on the occupation fraction of SMBHs in halos and the fraction of SMBHs hosting an accretion disc will also be well understood. Then, the only free parameters in the calculation of the DW background are the torque fluctuation model and the residence time ($\sim 1/\dot{f}$). In this case, the amplitude, slope and inflection points of the DW background become a probe of typical densities and temperatures of accretion discs, across all mass ranges and redshifts. These measurements could be used to calibrate and refine numerical simulations, as well as determine what disc models are more prevalent in the Universe.

In the second scenario, we imagine that the fluctuation spectra determined by hydrodynamical simulations turn out to be sufficiently generic and simple to be applicable to a wide range of masses and densities. Then, the DW background can become a tool to constrain statistical parameters in the population model. As an example, one could parametrize a DW background prediction with an arbitrary redshift-dependent fraction of SMBHs hosting an accretion disc and find a best-fitting model to the actual data. Such constraints are interesting, because they are completely orthogonal to other methods that require some electromagnetic measurements.

\subsubsection{Multiband sources}

A massive binary of the type discussed in Section~\ref{sec:Detectability:individual} presents the opportunity for a multiband detection of GWs (namely PTAs and LISA), in addition to the array of pre-existing electromagnetic observations. For example, a detection of OJ 287 as a nHz GW source would confirm the existence of a secondary SMBH. A subsequent detection of DWs in the mHz band would constrain the slope and amplitude of the high-frequency end of torque fluctuations on the secondary, yielding information on the density and temperature profiles of the disc, as well as the small-scale gas flow around the companion to OJ 287. 
Such a ``multiband-multimessenger'' source would be a valuable probe for the physics of quasar accretion.
 
\subsection{A note on other future gravitational wave detectors}\label{sec:Discussion:other}

Until this point, we have never commented on whether DWs and their effects might be more relevant for detectors that cover parts of the GW frequency space other than the mHz band. As an example, the sensitivity band of the proposed $\mu$Hz detector $\mu$-Ares \citep[][]{muares} would be perfectly suited to detect both the DW background and the DW strain from the gas-embedded PTA sources discussed in Section~\ref{sec:Detectability}. In the same vein, the proposed Japanese detector DECIGO \citep[][]{decigo} and future iterations of LIGO \citep[][]{advLIGO} could be sensitive enough to detect the DW strain (or background) that originates from gas-embedded LISA sources. Our detectability analysis is focused on LISA since it is currently in development. However, DWs are not limited to the mHz and could be detected with other future GW detectors. The calculations in this paper can be reworked to suit any planned detector whenever its construction becomes more concrete. Note also that, having a sensitivity strain curve that is very similar to LISA, our results will also likely apply to the GW detector TianQin \citep[see, e.g.][]{Tian2019,Tian2021}.

\section{Conclusion}\label{sec:Discussion:conclusion}

In this paper we introduced the idea that hydrodynamical stochastic torque fluctuations can leave an imprint in the GW output of gas-embedded binaries, even if their mean value vanishes when orbit-averaged. We found that the orbital buffeting caused by torque fluctuations produces small-amplitude high-frequency perturbations that ride on the main carrier GW, which we dub \textit{dirty waveforms} (DWs). We parametrized the unresolved power spectrum of sub-orbital fluctuations with a normalisation amplitude $\sigma$ and a spectral index $j$. We found several potential observational prospects of DWs, and investigated the detectability of these new environmental effects for future space-borne mHz GW detectors.

\begin{itemize}

\item We showed that the additional energy emitted by DWs can produce a dephasing for typical LISA sources (Eqs~\ref{eq:dephasingTI} and \ref{eq:dephasingTII}), in addition to the dephasing caused by linear gas torques. If identified, it could be used to constrain the density and temperature profiles of the accretion disc.

\item We showed that the DWs of future PTA single-source targets could shine all the way up into the LISA band, producing a nHz-mHz multiband GW source. If identified, their slope and amplitude could help constrain the density, temperature and gas flow properties of quasar accretion discs.

\item We showed that the sum of all DWs from gas-embedded SMBH mergers can produce a stochastic background in the mHz band. If identified, it could help constrain parameters such as the abundance of AGN hosting an SMBH binary and the prevalence of different disc models.
\end{itemize}

All of the above effects will be significant if the amplitude of high-frequency torque fluctuations persists beyond the limits of current hydrodynamical simulations. In our parametrization of the fluctuation spectrum, a spectral index of the order $j \lesssim 1/2$ is required for the effects to be visible with LISA. Of course, more speculative GW detectors such as $\mu$-Ares and DECIGO might be able to detect weaker effects.

To conclude, whether DWs and their observational signatures will produce observable effects for LISA depends on the small scale physics of accretion discs that host an intermediate or extreme mass-ratio binary. Currently, the detectability of DWs is speculative because they originate at a scale that can only be probed with sufficiently resolved 3D magnetohydrodynamical simulations. However, these limitations are mostly computational and therefore likely to improve significantly by the time LISA data will become available.

If DWs are loud enough, they will present an opportunity to use LISA as a $\textit{sophisticated probe}$ of accretion disc physics. Moreover, they will imply the existence of a new type of multi-band GW source, potentially visible with PTAs and LISA at the same time. If not, the ideas in this paper are still illustrative of a new type of environmental effect affecting the GWs of merging BH binaries, and are suggestive of the potential of actively searching for such traces in GW signals. We believe that an effort in identifying, quantifying, and adding environmental effects in waveform template searches is crucial to assure the success of future GW observatories. Now, half a dozen years after their first detection, it seems that we are just on the verge of being able to do $\textit{astronomy}$ with GWs.

\section{Data availability statement}
Simulation data referred to in this study (from \citealt{Derdzinski2021}) is available upon request from AD. 

\section*{Acknowledgements}
PRC, LM, and LZ acknowledge support from the Swiss National Science Foundation under the Grant 200020\_178949. AD acknowledges support from the Tomalla Foundation for Gravity Research. LZ also acknowledges the BAC for insightful discussions. MG acknowledges support from the Swiss National Science Foundation under the Grant 200020\_192092.

\scalefont{0.94}
\setlength{\bibhang}{1.6em}
\setlength\labelwidth{0.0em}
\bibliographystyle{mnras}
\bibliography{DW}
\normalsize


\bsp 
\label{lastpage}
\end{document}